\documentclass[aps,pra,showpacs,superscriptaddress,numerical,twocolumn,amsmath,amssymb,floatfix,reprint]{revtex4-1}

\usepackage[per-mode=symbol,separate-uncertainty]{siunitx}
\usepackage[english]{babel}
\usepackage{graphicx}
\usepackage{dcolumn}
\usepackage{bm}
\usepackage[protrusion=true, expansion=true]{microtype} 
\usepackage[
	pdffitwindow=true,
	colorlinks=true,
	frenchlinks=false,
    linkcolor=blue,
    anchorcolor=blue,
    citecolor=blue,
    filecolor=blue,
    urlcolor=blue,
    bookmarks=true,
    bookmarksopen=true,
	bookmarksnumbered=true,
    bookmarksopenlevel=1,
    plainpages=false,
	pdfpagelayout=SinglePage,
    pdfpagelabels=true,
	breaklinks
]{hyperref}
\usepackage[caption=false]{subfig}
\usepackage{braket}
\usepackage{multirow}
\usepackage{longtable}
\usepackage{ulem} 
\renewcommand\bra[1]{{\langle{#1}|}} 
\makeatletter
\renewcommand\ket[1]{%
  \@ifnextchar\bra{\k@t{#1}\!}{\k@t{#1}}%
}
\newcommand\k@t[1]{{|{#1}\rangle}}
\renewcommand{\onlinecite}{\cite}
\makeatother

\graphicspath{{figures/}}

\usepackage{verbatim}

\begin{document}

\preprint{APS/123-QED}

\title{Tunable refrigerator for non-linear quantum electric circuits}
\author{Hao Hsu}
\affiliation{JARA Institute for Quantum Information (PGI-11), Forschungszentrum J{\"u}lich, 52425 J{\"u}lich, Germany}
\author{Matti Silveri}
\affiliation{QCD Labs, QTF Centre of Excellence, Department of Applied Physics, Aalto University, P.O.~Box 13500, FI-00076 Aalto, Finland}
\affiliation{Research Unit of Nano and Molecular Systems, University of Oulu, P.O.~Box 3000, FI-90014 Oulu, Finland}
\author{Andr\'as Gunyh\'o}
\affiliation{QCD Labs, QTF Centre of Excellence, Department of Applied Physics, Aalto University, P.O.~Box 13500, FI-00076 Aalto, Finland}
\author{Jan Goetz}
\affiliation{QCD Labs, QTF Centre of Excellence, Department of Applied Physics, Aalto University, P.O.~Box 13500, FI-00076 Aalto, Finland}
\affiliation{IQM, Keilaranta 19, FI-02150 Espoo, Finland}
\author{Gianluigi Catelani}
\affiliation{JARA Institute for Quantum Information (PGI-11), Forschungszentrum J{\"u}lich, 52425 J{\"u}lich, Germany}
\author{Mikko M\"{o}tt\"{o}nen}
\affiliation{QCD Labs, QTF Centre of Excellence, Department of Applied Physics, Aalto University, P.O.~Box 13500, FI-00076 Aalto, Finland}
\affiliation{VTT Technical Research Centre of Finland Ltd., \\ P.O. Box 1000, FI-02044 VTT, Finland}
\date{\today}

\begin{abstract}

The emerging quantum technological applications call for fast and accurate initialization of the corresponding devices to low-entropy quantum states. To this end, we theoretically study a recently demonstrated quantum-circuit refrigerator in the case of non-linear quantum electric circuits such as superconducting qubits.
The maximum refrigeration rate of transmon and flux qubits is observed to be roughly an order of magnitude higher than that of usual linear resonators, increasing flexibility in the design.
We find that for typical experimental parameters, the refrigerator is suitable for resetting different qubit types to fidelities above 99.99\% in a few or a few tens of nanoseconds depending on the scenario.
Thus the refrigerator appears to be a promising tool for quantum technology and for detailed studies of open quantum systems.
\end{abstract}
\maketitle

\section{Introduction}\label{sec:intro}
Superconducting quantum circuits have proved to be promising building blocks for future quantum computing~\cite{Versluis_2017,Heinsoo_2018,Rosenblum_2018,Neill_2018}, quantum simulation~\cite{Forn-Diaz_2016,Braumueller_2017,Langford_2017,Goetz_2018}, and sensing~\cite{Govenius_2016,Goetz_2016a,Inomata_2016,Besse_2018,Kono_2018} applications. Importantly, a superconducting quantum computer has been reported to achieve quantum supremacy~\cite{Arute_2019} which is a stepping stone to quantum advantage and ultimately to a fault-tolerant error-corrected quantum computer~\cite{Preskill_2018}.

Precise initialization is a key requirement for the reliable operation of superconducting circuits in their  quantum-technological applications. This is especially pronounced in multi-qubit quantum processors, where the initialization fidelity for the whole quantum register decreases exponentially with the number of qubits assuming a fixed fidelity for single-qubit initialization.

The simplest approach to initialize quantum circuits is to cool the surrounding environment to millikelvin temperatures using a dilution refrigerator and to wait for a spontaneous decay. However, since the relevant coherence times are approaching the millisecond regime~\cite{Rigetti_2012,Riste_2013,Yan16}, this method becomes impractical in general due to the long waiting time. In non-error-corrected processors, the waiting time typically only reduces the repetition rate of the experiment. However, if many ancilla qubits are needed in the executed algorithm, an initialization scheme operating much faster than the natural decoherence of the qubits may greatly save resources. Namely, one could reuse individual physical qubits many times as ancillas in the same algorithmic run without errors being accumulated into them. For quantum error correction, the need for such fast reset is especially important since a large number of ancilla qubits is needed during every quantum-error-correction cycle~\cite{Terhal_2015,Fowler_2009,Fowler_2011,Nigg_2014}. Thus in practice, active reset is the preferred path to the fault-tolerant quantum computer.

Active reset protocols have been developed based on frequency tuning~\cite{Reed_2010}, feedback control~\cite{Riste_2012,Johnson_2012,Campagne_2013,Salathe_2018} and microwave driving~\cite{Valenzuela_2006,Grajcar_2008,Geerlings_2013a,McClure_2016,Magnard_2018,Egger_2018}. For example, Ref.~\cite{Magnard_2018} reports initialization of the qubit-resonator system with~\SI{0.2}{\percent}~error to its ground state in less than~\SI{500}{\nano\second}, an order of magnitude faster than the natural decoherence of the utilized qubit.
In an effort to yet improve the initialization speed and fidelity, tunable dissipators have been proposed~\cite{Jones_2013,Tuorila_2017,Schreier08,Tuorila_2019} and experimentally tested~\cite{Partanen_2018,Partanen_2019}. Since the qubit state is here modified only through its environment, these dissipators have the potential to prevent initialization errors arising from imprecise qubit control and readout and from non-adiabatic excitations owing to the on/off modulation of the dissipation.

However, the above-discussed active reset schemes have the disadvantage that they only empty the first or a few of the lowest excited states in the circuits. Thus leakage of quantum information out of the computational space may accumulate and pose a bottleneck to fidelity.

Fortunately, recent work has given rise to the concept of a quantum-circuit refrigerator (QCR)~\cite{Tan_2016}, which provides a broadband voltage-tunable environment for quantum electric circuits, thus being able to dissipate also highly excited states of the circuits.
The QCR consists of a superconductor--insulator--normal-metal--insulator--superconductor~(SINIS) junction in which the normal part is capacitively coupled to the circuit of interest. The rate of electron tunneling through the junction, and thus of energy exchange between QCR and circuit, can be controlled by a bias voltage.
In the case of cooling a superconducting resonator, its operational principles have been theoretically studied, starting from a microscopic model~\cite{Silveri_2017}. In the experiments of Ref.~\cite{Silveri_2019}, this model was found to describe accurately both the tunability of the resonator dissipation rate by orders of magnitude and the consequently induced Lamb shift of the resonance frequency. Reference~\cite{Sevriuk_2019} experimentally demonstrated that the QCR can be turned on/off in nanosecond time scale, which seems adequate for qubit initialization.

In this paper, we extend the previous theory limited to linear resonators to non-linear quantum electric circuits typically used as superconducting qubits. We find that independent of the qubit type, transmon or capacitively shunted flux qubit, it seems experimentally feasible to integrate the QCR to the quantum processor for improved initialization. Namely, the residual decay rate, the dephasing rate, and the Lamb shift introduced by the QCR can be engineered to be modest while retaining the fast initialization speed. We also show that the introduction of a readout resonator to the QCR-coupled qubit does not introduce a major challenge. In fact, the QCR may also be efficiently used to initialize the  resonator if needed. Thanks to the high tunability and versatility of the QCR, our results may pave the way for deeper understanding of open quantum systems beyond the usual Born--Markov and secular approximations~\cite{Tuorila_2019}.

This paper is organized as follows: Section~\ref{sec:circuitcoupling} provides a theoretical description of a QCR coupled to a general quantum electric circuit. In Sec.~\ref{sec:operation}, we discuss the on/off operation regimes and the effective temperature of the QCR environment which are independent of the qubit type. In Sec.~\ref{sec:qubits}, we present the transition rates that the QCR induces on a transmon and on a capacitively shunted flux qubit at the on/off operation regimes. Section~\ref{sec:dephasing} is devoted to quantifying the dephasing introduced by the QCR on the qubits. Section~\ref{sec:Lamb} provides a compact equation to compute the Lamb and ac Stark shifts of the different transitions in the non-linear circuit coupled to the QCR. Section~\ref{sec:QCR_tr_res} takes a step to an even higher circuit complexity by considering a QCR coupled to a system composed of a resonator and a transmon qubit. Section~\ref{sec:conclusions} concludes this paper and the appendices provide additional details of the derivations.

\section{Coupling the refrigerator to an arbitrary quantum electric circuit}
\label{sec:circuitcoupling}

\begin{figure}[tb]
    \centering
    \includegraphics[width=1\linewidth]{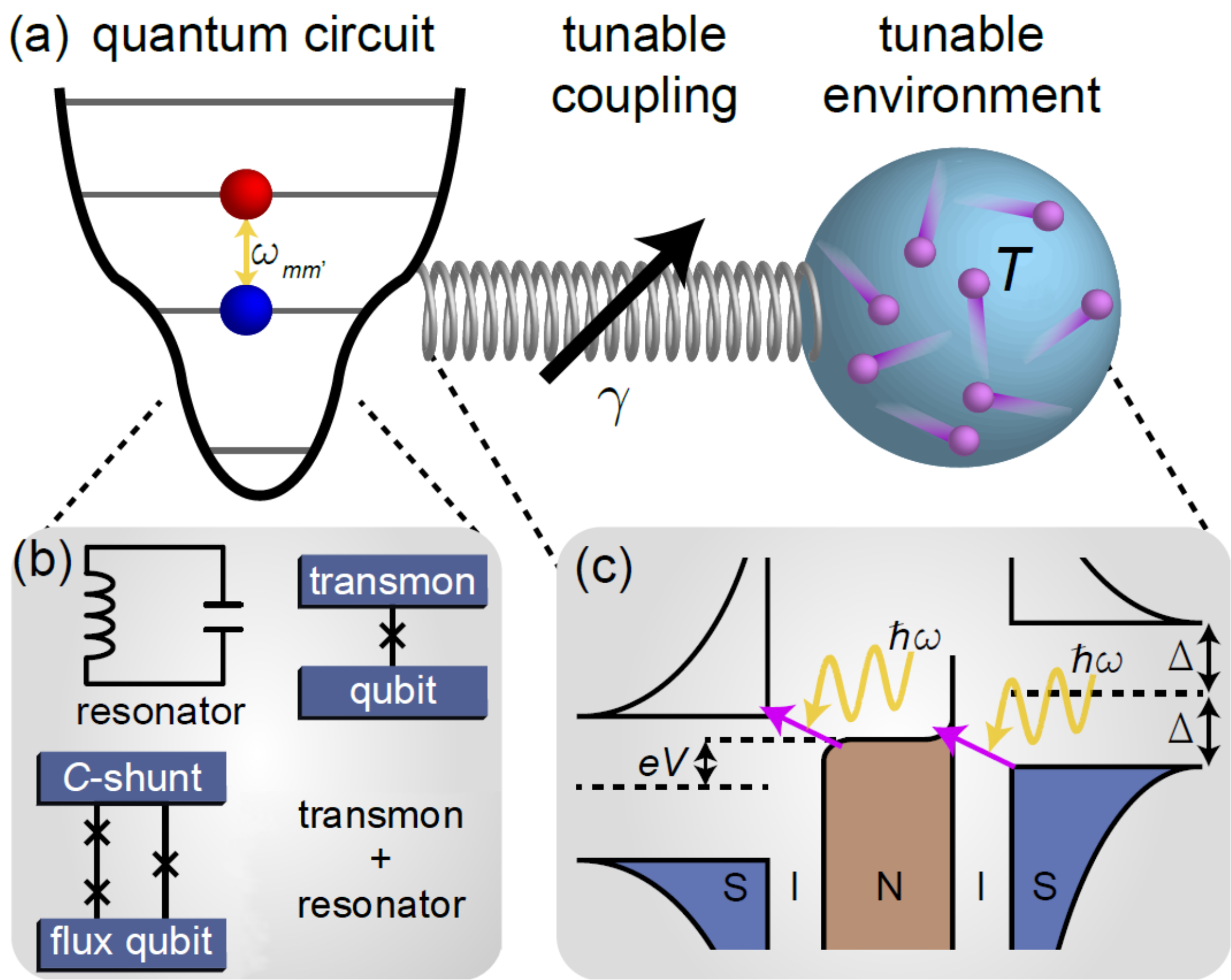}
    \caption{\label{fig:01} (a) Quantum circuit coupled with a tunable coupling strength $\gamma$ to a broadband environment at an effective temperature $T$. We consider an arbitrary potential landscape of the quantum circuit resulting in energy levels $E_{m}$ separated by the transition energies $\hbar\omega_{mm^{\prime}}\,{=}\,E_{m}\,{-}\,E_{m^{\prime}}$. (b)~The quantum circuit can for example be a resonator, a superconducting qubit or any combination thereof. (c) The tunable environment is realized by photon-assisted electron tunneling through a superconductor--insulator--normal-metal--insulator--superconductor~(SINIS) junction. Tunneling is controlled by the bias voltage~$V$. We refer to the bias-voltage-controlled and capacitively coupled SINIS junction as the quantum-circuit refrigerator~(QCR). A circuit diagram for the coupled system is presented in Fig.~\ref{fig:QCR_circuit_gen}.}
\end{figure}

As shown in Fig.~\ref{fig:01}(a), we consider a quantum electric circuit coupled to a broadband environment. Our theory is not restricted to any particular circuit type but it can be applied to any kind of electrical circuit such as superconducting qubits, resonators, or combinations thereof, see Fig.\,\ref{fig:01}(b). We only assume here that the quantum circuit has discrete energy levels $E_{m}$. The broadband environment at the effective temperature $T$ is realized by tunneling electrons that couple to the quantum circuit through photon absorption and emission, see Fig.\,\ref{fig:01}(c). One way of interpreting this circuit structure is within the spin-boson model~\cite{Leggett_1987} where only two isolated energy levels, i.e., a spin-1/2, is coupled with a coupling strength $\gamma$ to a bosonic bath. However, since the environment is provided by fermionic degrees of freedom and the circuit is a multilevel device, we cannot precisely describe the system within the spin-boson model. Instead, we show below how the tunable coupling strength $\gamma(V)$ and the tunable effective temperature $T(V)$ can be derived from electron tunneling through a pair of normal-metal–-insulator--superconductor (NIS) junctions biased by a voltage $V$.

Let us discuss how a quantum circuit interacts with a QCR. The QCR is coupled through a capacitance $C_c$ to the quantum circuit,
the excitations of which 
are absorbed by electrons tunneling across the NIS junctions
at a  rate which depends on the voltage $V$.
Owing to the capacitive coupling, the tunneling leads to a charge offset in the quantum circuit. We denote the matrix element between eigenstates $\ket{m}$ and $\ket{m'}$ of the circuit due to this charge offset by $M_{mm'}$, the exact definition of which is given in Eq.~\eqref{eqn:Mmm} below.
The case where the circuit is a resonator is studied in Ref.~\onlinecite{Silveri_2017} where using Fermi's golden rule it is shown that the transition rates are given by
\begin{equation}
    \Gamma_{mm^{\prime}}(V) = \left|M_{mm^{\prime}}\right|^{2}\frac{2R_{K}}{R_T}\sum_{\tau=\pm1}F(\tau eV + \hbar\omega_{mm'} - E_{N}),
    \label{eqn:Gamma}
\end{equation}
where $R_{K}\,{=}\,h/e^{2}=25.813\textrm{ k}\Omega$ is the von Klitzing constant, $R_{T}$ is the tunneling resistance of the NIS junctions, $\hbar\omega_{mm'} = E_{m} - E_{m'}$ denotes the transition energy between the quantum circuit eigenstates $\ket{k}$ at the energies $E_k$, and $E_{N}\,{=}\,e^{2}/(2C_{N})$ is the charging energy of the normal-metal island. In this manuscript, we denote the elementary charge by $e$, the Planck constant by $h$, and $\hbar=h/(2\pi)$.
The function $F$  describes the normalized rate of single electron tunneling,
\begin{equation}
F(E) = \int\!d\varepsilon~\frac{n_{S}(\varepsilon)}{h} \frac{f (\varepsilon-E) -f (\varepsilon)}{1-e^{-E/(k_B T_N)}}\,, \label{eqn:F}
\end{equation}
where $n_S$ is the normalized superconductor density of states. Above, we assume a thermal equilibrium at temperature $T_N$ for the quasiparticle excitations in both the normal metal and the superconductor, characterized by the Fermi-Dirac distribution function $f(\varepsilon)$ at $T_N$. 
We define the Dynes parameter $\gamma_D$ according to
\begin{equation}
n_{S}(\varepsilon) = \left|\mathrm{Re} \left[ \frac{\varepsilon + i \gamma_{D}\Delta}{\sqrt{(\varepsilon + i \gamma_{D}\Delta)^{2} - \Delta^{2}}}\right] \right|,
\label{eqn:ns}
\end{equation}
thus accounting for the effective broadening of density of states from various different sources. Analytical approximations for $F(E)$ in different relevant parameter regimes can be found in Appendix~\ref{app:F_E}.

\begin{figure}[tb]
    \includegraphics[width=1\linewidth]{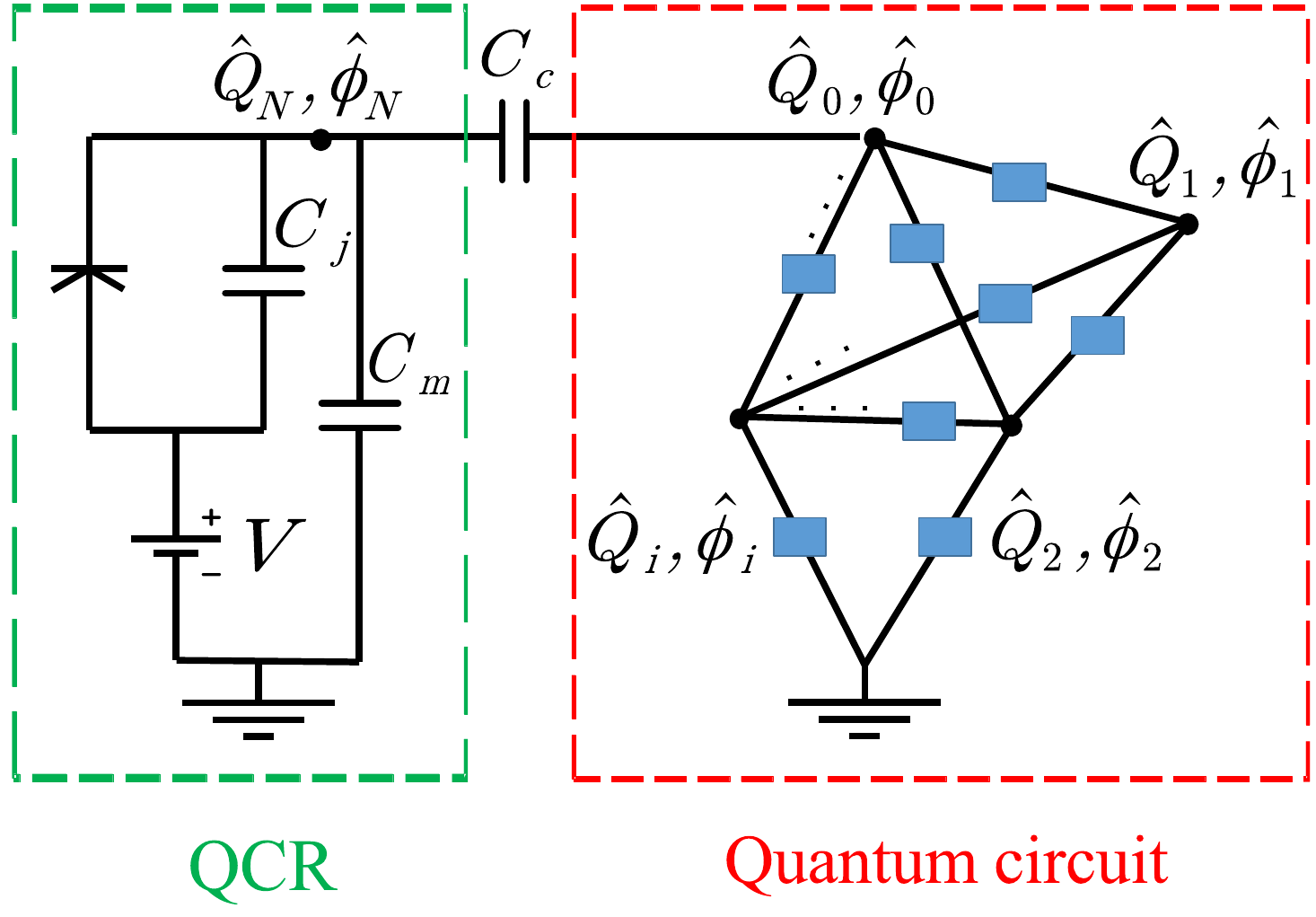}
    \caption{\label{fig:QCR_circuit_gen} Effective circuit diagram for a quantum-circuit refrigerator (QCR) capacitively coupled to an arbitrary quantum-circuit network. In the network, a blue box connecting a pair of nodes denotes a capacitor, an inductor, a Josephson junction, or parallel combinations thereof. The voltage-biased SINIS junction operating as the QCR is effectively described by the circuit in the green dashed box.}
\end{figure}

Next, we show that the transition rates of Eq.~\eqref{eqn:Gamma} can be generalized to an arbitrary quantum circuit. As depicted in Fig.~\ref{fig:QCR_circuit_gen}, we denote with $\Phi_0$ and $Q_0$ the flux and charge at the node of the circuit coupled to the QCR, respectively. The circuit itself is an arbitrary network of $M+1$ nodes connected possibly by capacitors, inductors, and Josephson junctions. We write the Hamiltonian $\hat H_\mathrm{tot}$ for the total system as
\begin{align}
\hat H_\mathrm{tot} =& \hat H_0 + \hat H_N + \hat H_S + \hat H_T,  \label{eqn:Htot}
\end{align}
where $\hat H_N$, $\hat H_S$, and $\hat H_T$ are the microscopic Hamiltonians describing the normal metal, superconductor, and their tunnel coupling, respectively, and the core Hamiltonian
\begin{align}
\hat H_0  =& \hat H_C + \hat H_\phi, \label{eqn:H0}
\end{align}
 accounts for the quantum circuit, the QCR charging energy, and the QCR--circuit capacitive coupling. As shown in Appendix~\ref{app:inverseC}, the total charging energy part $\hat H_C$ assumes the form
\begin{align}
&\hat H_C = \frac{1}{2C_N}\hat Q_N^2
+  \frac{1}{2\widetilde{C}_0}\left(\hat Q_0+\alpha \hat Q_N\right)^2  \label{eqn:HC} \\ &+\left(\hat Q_0+\alpha  \hat Q_N\right) \sum_{i=1}^M \left(\widetilde{C}_V^{-1}\right)_{0i} \hat Q_i + \frac12\!\sum_{i,j=1}^M\left(\widetilde{\boldsymbol{C}}^{-1}_M\right)_{ij} \hat Q_i \hat Q_j, \notag
\end{align}
where the capacitance ratio
\begin{equation}\label{eqn:alpha}
\alpha = C_c/C_N,
\end{equation}
plays the role of a dimensionless coupling constant ($0<\alpha<1$) with the coupling capacitance $C_c$, total normal-metal capacitance $C_N = C_c + C_\Sigma$, and the total junction capacitance $C_\Sigma= C_j + C_m$, see Fig.~\ref{fig:QCR_circuit_gen} for definitions and Table~\ref{tab:parameters} for typical values. In this form, the charge offset caused by the QCR on the connected node becomes evident. The inverse capacitance vector $\widetilde{C}_V^{-1}$ and matrix $\widetilde{\boldsymbol{C}}_M^{-1}$ are related to the circuit capacitance matrix and the coupling capacitance $C_c$ as defined explicitly in Appendix~\ref{app:inverseC}; they, as well as the potential energy part $\hat H_\phi$, depend on the details of the circuit which we will consider in some concrete examples in Secs.~\ref{sec:qubits}~and~\ref{sec:QCR_tr_res}.

\renewcommand{\arraystretch}{1.25}
\begin{table}[bt]
    \centering
    \begin{tabular}{c c c c c c c}
         \multicolumn{1}{c|}{} &  $C_c$ & $C_N$ & $R_T$ & $\gamma_D$ & $\Delta$ \\
         \cline{1-6}
         \multicolumn{1}{c|}{QCR} & \SI{1}{\pico\farad} & \SI{1.01}{\pico\farad} & \SI{50}{\kilo\ohm} & \num{e-5} & \SI{200}{\micro\electronvolt} \\
         \\[-0.2cm]
        \multicolumn{1}{c|}{} & $\omega_{10}/2\pi$ & $E_{J}/E_C$ \\
         \cline{1-3}
        \multicolumn{1}{c|}{Transmon} & \SI{5}{\giga\hertz} & \num{50} \\
         \\[-0.2cm]
         \multicolumn{1}{c|}{} & $\omega_{10}/2\pi$ & $E_{J}/E_C$ & $E_{J}/h$ & $\alpha_\mathrm{fl}$& $\zeta$ & $f_e$\\
         \cline{1-7}
         \multicolumn{1}{c|}{\vtop{\hbox{\strut C-shunted}\hbox{\strut flux qubit}}} & \multirow{2}{*}{\SI{4.1}{\giga\hertz}} & \multirow{2}{*}{\num{103.4} } & \multirow{2}{*}{ \SI{86.2}{\giga\hertz} } & \multirow{2}{*}{\num{0.42} } & \multirow{2}{*}{\num{2.2}} &
         \multirow{2}{*}{\num{0.5}}

    \end{tabular}
    \caption{Parameter values of the devices. The QCR parameter values are as in Ref.~\onlinecite{Silveri_2017}. The transmon parameters (qubit frequency and ratio between Josephson and charging energy) agree to those of state-of-the-art transmons~\cite{Paik11, Rigetti_2012, Chang_2013}. For the capacitively shunted flux qubit, the parameters (with $\alpha_\mathrm{fl}$ characterizing the small junction, $\zeta$ the shunting capacitor, and $f_e$ the flux frustration, see Sec.~\ref{sec:CSFQ}) are identical to those of Ref.~\onlinecite{Yan16} (sample B).}
    \label{tab:parameters}
\end{table}

The Hamiltonians $\hat H_N$ and $\hat H_S$ account for the energy of excitations in the normal-metal and in the superconductor leads in the NIS junction of the QCR and the tunneling across this junction is described by the Hamiltonian~$\hat H_T$~\cite{Silveri_2017}:
\begin{align}
\hat H_N = & \sum\limits_{l\sigma} \varepsilon_l \hat d^\dagger_{l\sigma} \hat d^{}_{l\sigma}, \\
\hat H_S = & \sum\limits_{k\sigma} \epsilon_k \hat c^\dagger_{k\sigma} \hat c^{}_{k\sigma}+ \sum\limits_{k} (\tilde \Delta_k \hat c^\dagger_{k\uparrow} \hat c^\dagger_{-k\downarrow} + \textrm{H.c.} ), \\
\hat H_T = & ~\hat \Theta e^{-i\frac{e}{\hbar}\left(\hat \phi_N -  Vt\right)} + \mathrm{H.c.} \, , \quad \hat \Theta = \sum_{kl\sigma} T_{lk} \hat d^\dagger_{l\sigma} \hat c^{}_{k\sigma}, \label{eq:HT}
\end{align}
where $\varepsilon_l$ and $\epsilon_k$ denote the excitation energies in the normal metal and in the superconductor, respectively, $\hat d_{l\sigma}$ and $\hat c_{k\sigma}$ are the corresponding annihilation operators, and $\tilde \Delta_k=\Delta_k e^{-i \frac{2e}{\hbar}Vt}$ is the superconductor gap parameter after a time-dependent unitary transformation $\hat U_V (t)=\Pi_{k\sigma }\exp(i\frac{e}{\hbar}Vt \hat c_{k\sigma }^\dag  \hat c^{}_{k\sigma})$  for transforming the effect of the voltage bias $V$ to the operators. Note that the total Hamiltonian transforms according to $\hat H''_\textrm{tot}= \hat U_V^\dag \hat H_\textrm{tot}  \hat U^{}_V + i\hbar (\partial _t  \hat U_V^\dag)  \hat U^{}_V$. 
The tunneling matrix elements $T_{lk}$ provide the amplitude of the corresponding tunneling event.

We carry out the transformation $\hat H_C'  = \hat U \hat H_C \hat U^\dagger$ by the unitary operator $\hat U= e^{\frac{i}{\hbar}\alpha \hat Q_N \hat \phi_0}$ that removes the charge offset from the capacitive energy Hamiltonian $\hat H_C$ such that
\begin{align}
\hat H_C'  = & \frac{1}{2C_N}\hat Q_N^2  + \frac{1}{2\widetilde{C}_0}\hat Q_0^2 \label{eqn:HCp} \\ & + \hat Q_0 \sum_{i=1}^M \left(\widetilde{C}_V^{-1}\right)_{0i} \hat Q_i + \frac12\sum_{i,j=1}^M\left(\widetilde{C}^{-1}_M\right)_{ij} \hat Q_i \hat Q_j. \notag
\end{align}
All other terms in the total Hamiltonian $\hat H_\mathrm{tot}$ are left unaffected, except for the tunneling part $\hat H_T$ which becomes
\begin{equation}
\hat H_T' = \hat \Theta e^{-i\frac{e}{\hbar}\left(\hat \phi_N - Vt\right)} e^{-i\frac{e}{\hbar} \alpha \hat \phi_0} + \mathrm{H.c.}.
\label{eqn:HTp}
\end{equation}
Treating the tunneling Hamiltonian $\hat H_T'$ as a perturbation, the states $|E,q,m\rangle$ of the unperturbed Hamiltonian can be labeled by the total energy $E$ of the quasiparticle excitations in the normal metal $N$ and in the superconductor $S$ leads, by the charge on the normal-metal island $q$, and by the state index of the quantum circuit $m$. The calculation of the transition rates due to $\hat H_T'$ proceeds as explained in Ref.~\onlinecite{Silveri_2017}: the factors $e^{\mp i\frac{e}{\hbar}\hat \phi_N}$ force the final charge on the normal-metal island $q'$ to satisfy $q'=q\pm 1$, and after averaging over the initial states of the quasiparticle excitations, we find the transition rate of Eq.~\eqref{eqn:Gamma} where the matrix elements are given by
\begin{equation}\label{eqn:Mmm}
M_{mm'}= \bra{m'} e^{-i\frac{e}{\hbar} \alpha \hat\phi_0} \ket{m}.
\end{equation}
We stress that to arrive at Eq.~\eqref{eqn:Gamma}, a number of simplifying assumptions have been made: the elastic transition rates $\Gamma_{mm}$ are assumed to be fast compared to the other rates; this implies that the QCR charge states quickly reach a steady state, and therefore we may calculate the transition rates of the circuit by averaging over the charge distribution. In fact, it was shown in Ref.~\onlinecite{Silveri_2017} that the distribution is thermal-like, but with an effective temperature which differs from the normal-metal electron temperature $T_N$ at finite bias. To obtain this result, it was also assumed that the charging energy $E_N$ is the smallest energy scale, i.e., $E_N \ll k_B T_N, \hbar \omega_{mm'}$ for $m\neq m'$. Naturally, the relevant temperatures and frequencies are smaller than the superconductor gap parameter $\Delta$.

Note that the procedure described above to express the matrix elements in the form of Eq.~\eqref{eqn:Mmm} is in practice suitable if $\phi_0$ is not a compact variable, that is, if an inductor is connected to node~\num{0} of the network of Fig.~\ref{fig:QCR_circuit_gen}. In this case, we have $\langle \phi_0|m\rangle\to 0$ for $\phi_0\to \pm\infty$ and any $\ket{m}$, where $\ket{\phi_0}$ is the eigenstate of the operator $\hat\phi_0$ with the eigenvalue $\phi_0$. 
If in the original formulation of Eqs.~\eqref{eqn:Htot}--\eqref{eqn:HC} the wavefunction is periodic in $\phi_0$, then after the unitary transformation in Eq.~\eqref{eqn:HCp}, we should impose twisted boundary conditions dependent on the charge state $q$. In such a situation it is more practical not to introduce the unitary transformation and to obtain the matrix element from the overlap of the relevant eigenstates, $M_{mm'} = \langle m'_{q+1} \ket{m_q} = \langle m'_{1} \ket{m_0}$. In practice, we numerically diagonalize the core Hamiltonian $\hat H_0$ for offset charges $Q_N=-\alpha eq$ with $q=0$ and $1$, and then numerically calculate the overlap between the obtained eigenstates with $q=0$ and $1$. We adopt this method to calculate the matrix elements $M_{mm'}$ in Sec.~\ref{sec:qubits} for different kinds of 
weakly anharmonic superconducting qubits. However, the theory developed throughout this section is general and can be applied to systems of arbitrary anharmonicity.

\section{Operation regimes}\label{sec:operation}

\begin{figure*}
  \centering
    \includegraphics[width=1\linewidth]{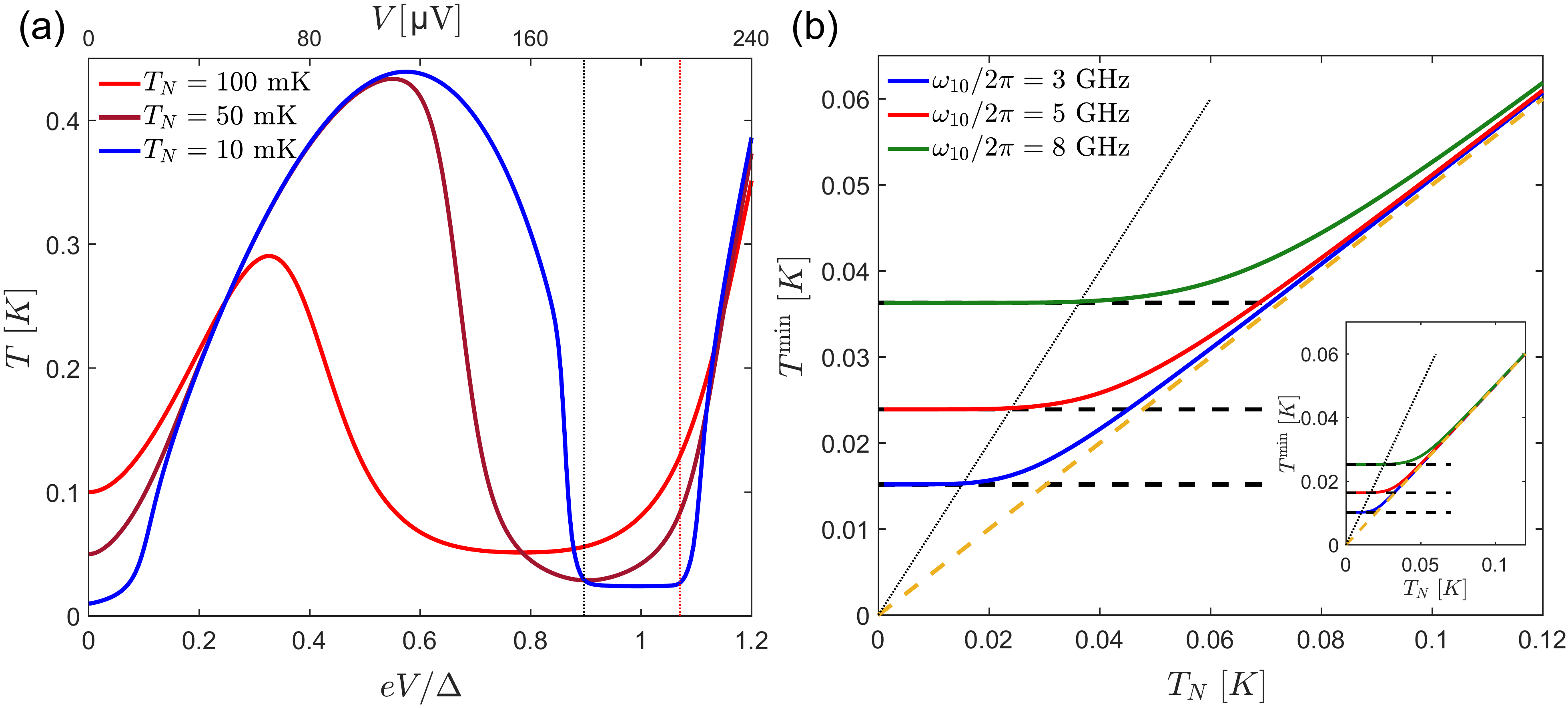}
    \caption{\label{fig:02} (a) Effective temperature $T$ of the environment as a function of the single-junction bias voltage $V$ for three values of the normal-metal electron temperature $T_N=$~\SI{100}{\milli\kelvin}~(bright red), \SI{50}{\milli\kelvin}~(dark red),~\SI{10}{\milli\kelvin}~(blue). The qubit transition frequency is $\omega_{10}/(2\pi)=$~\SI{5}{\giga\hertz}. The vertical dotted lines denote the on-voltage in the thermal activation regime $eV=\Delta-\hbar\omega_{10}$ (black) and the on-voltage of Eq.~\eqref{eq:eVmax} in the low-temperature regime $eV=eV_\mathrm{max}$ for $T_N=$~\SI{10}{\milli\kelvin} (red).
    The top horizontal scale in this and subsequent plots gives the bias voltage based on the typical gap value for thin Al films, see Table~\ref{tab:parameters}. (b) Minimum effective temperature
    as a function of the normal-metal electron temperature $T_{N}$ for three values of the qubit transition frequency $\omega_{10}/2\pi=$~\SI{3}{\giga\hertz} (blue), \SI{5}{\giga\hertz} (red), and \SI{8}{\giga\hertz} (green). The solid lines are results calculated by numerically minimizing  Eq.~\eqref{eqn:T_T} in the on-state, whereas the horizontal black dashed lines are calculated using Eq.~\eqref{eqn:T^min_T}. For comparison, the diagonal orange dashed and black dotted lines denote $T^{\mathrm{min}}= T_{N}/2$ and $T^{\mathrm{min}}=T_{N}$, respectively. In the main figures, the Dynes parameter is $\gamma_{D} =$~\num{e-5}, but in the inset $\gamma_{D} =$~\num{e-7}. The other quantum-circuit parameters are given in Table~\ref{tab:parameters}.
    }
\end{figure*}

In this section, we define and characterize the off and on operation regimes. The off-state is defined by bias-voltage values, for which the effect of the QCR on the quantum circuit is negligible or the weakest possible. In the on-state, the bias voltage is such that the QCR causes transitions in the quantum circuit with high rates. For simplicity and for practical relevance, we restrict our attention to the two lowest energy levels of a quantum circuit, that is, we assume that the quantum circuit is operated as a qubit  with the energy difference $\hbar\omega_{10}$.  As discussed in Sec.~\ref{sec:circuitcoupling}, we can characterize the tunable environment formed by the QCR using its coupling strength $\gamma(V)$ and effective temperature $T(V)$ which are related to the transition rates by
\begin{align}
\gamma(V) =& \Gamma_{10}-\Gamma_{01}, \label{eqn:gamma_T} \\
T(V) = & \frac{\hbar\omega_{10}}{k_{{B}}} \left[ \mathrm{ln} \left( \frac{\Gamma_{10}}{\Gamma_{01}} \right) \right]^{-1}. \label{eqn:T_T}
\end{align}
Inverting these definitions, we have~\cite{Silveri_2017}
\begin{equation}
\Gamma_{kl}(V) = \gamma(V)\left[N(V) + k\right],
\end{equation}
where $(k,l)\in\{(0,1), (1,0)\}$ and the effective mean occupation number is $N(V)=1/\left\{ \exp \left[\hbar\omega_{10}/(k_{{B}} T)\right]-1\right\}$.

\subsection{Off-state}
As visible in Fig.~\ref{fig:01}(c), there is an increasing number of tunneling channels available for an increasing bias voltage. Thus the QCR-induced transition rates are increasing functions of the bias voltage, and the off-state of the QCR is achieved at low bias voltages.

Since the matrix elements drop out from $T$ in Eq.~\eqref{eqn:T_T}, the effective temperature $T$ depends on the properties of the qubit only through its frequency. Therefore, we can study the effective temperature $T$ for a generic qubit, without considering a particular realization. In the limit $V\to 0$, the effective temperature approaches the electron temperature $T \to T_N$, as expected in thermal equilibrium~\cite{Silveri_2017}. However, the off-state is generically not limited to small bias, but defined in a wider, temperature-dependent range of voltage in which the effective temperature can significantly exceed $T_N$ as discussed in more detail in Appendix~\ref{app:onoff}. In the main text, we only consider for simplicity the off-state as that with $V=0$.

\subsection{On-state}
In the on-state, typically the bias~$eV$ is of the order of the superconducting gap parameter $\Delta$, and we distinguish two possibilities. If the normal-metal electron temperature $T_{N}$ of the QCR is higher than the cross-over value $T_{N}^{\rm co}$ defined below, then the broadening of the density of states characterized by the Dynes parameter $\gamma_{D}$ can be neglected. Effectively, one can take the limit $\gamma_D \to 0$ and recover the thermal activation regime of Ref.~\onlinecite{Silveri_2017}, in which the environment can reach an effective temperature below the equilibrium one, $T \approx T_{N}/2$, over a finite voltage range with  $V\lesssim(\Delta-\hbar\omega_{10})/e$, see the vertical black dotted line in Fig.~\ref{fig:02}(a). In fact, for $T_N>T_N^\text{co}$, we set $eV=\Delta-\hbar\omega_{10}$ as the on-state voltage, see Appendix~\ref{app:onoff} for details. This approximation, however, fails in the low-temperature regime $T_{N} < T_{N}^\mathrm{co}$. In this case, a good approximation is found if we take the limit $T_N = 0$, in which we can treat the distribution functions of Eq.~\eqref{eqn:F} as step functions and evaluate the integrals exactly. Thus, for $|E-\Delta| \gg \gamma_{D} \Delta$, we have approximately
\begin{equation}\label{eq:F_0T}
F(E) \approx \left\{ {\begin{array}{ll}
  \frac{1}{h} \sqrt{E^2-\Delta^2}, & E>\Delta, \\
  \frac{\gamma_{D} \Delta E}{h\sqrt{\Delta^2-E^2}}, & E<\Delta.  \\
\end{array}} \right.
\end{equation}
Using these expressions in Eq.~\eqref{eqn:Gamma} for the transition rates, substituting the results into the definition of the effective temperature of Eq.~\eqref{eqn:T_T}, and then minimizing the temperature with respect to the bias voltage $V$, we find that the minimum effective temperature at $T_{N} = 0$ is
\begin{equation}
T^{\mathrm{min}}(0) \simeq \frac{\hbar\omega_{10}}{k_{{B}}} \left\{ \ln \left[ \frac{2\hbar\omega_{10}}{\gamma_{{D}} \Delta (1-\frac{\hbar\omega_{10}}{\Delta})} \right] \right\}^{-1}. \label{eqn:T^min_T}
\end{equation}
The non-vanishing effective temperature at $T_N=0$ is reminiscent of the finite non-equilibrium shot noise at zero temperature, see Ref.~\onlinecite{Roussel16} and references therein. This minimum value is obtained for $eV \approx \Delta -(\hbar\omega_{10})^2/(2\Delta)$, but the effective temperature $T$ depends weakly on $eV$ over the finite tuning range approximately from $\Delta-\hbar\omega_{10}$ to
\begin{equation}
\label{eq:eVmax}
    eV_\mathrm{max}  = \Delta + \hbar\omega_{10} - e\delta V,
\end{equation}
where the voltage shift
\begin{equation}
    e\delta V  = k_B T_N\!\left\{\textrm{ln}\left( \frac{\sqrt{\pi} k_B T_N}{\gamma_D \Delta}\right) + \frac{1}{2} \textrm{ln}\left[ \textrm{ln} \left( \frac{\sqrt{\pi} k_B T_N}{\gamma_D \Delta} \right)\right]\!\right\} \label{eq:eVmax2}
\end{equation}
ensures that the excitation rates and hence the effective temperature do not start rising. For $T_N< T_N^\text{co}$, we thus choose the on-state bias voltage to be $V_\text{max}$, see the vertical red dotted line in Fig.~\ref{fig:02}(a) and Appendix~\ref{app:onoff}.

Note that the minimum effective temperature at $T_N=0$ in Eq.~(\ref{eqn:T^min_T}) depends only on the qubit frequency and on the superconductor properties of the QCR, $\Delta$ and $\gamma_D$, and it is small compared with frequency, $k_BT^\mathrm{min}(0) \ll \hbar\omega_{10}$.
When lowering the normal-metal electron temperature $T_{N}$, we approximate the cross-over from the thermal activation regime, where $T^{\mathrm{min}}\approx T_N/2$, to the low-temperature regime to occur at the temperature
\begin{equation}
    T_N^{\rm co} \approx 2 T^{\mathrm{min}}(0),
\end{equation}
see also Appendix~\ref{app:onoff}. This behavior is confirmed in Fig.~\ref{fig:02}(b) by numerical studies of the minimum effective temperature $T^{\mathrm{min}}$ as a function of the normal-metal electron temperature $T_{N}$ for different qubit frequencies and Dynes parameters. Note that as we decrease the electron temperature $T_{N}$, the effective temperature $T$ is bounded from below by $T^{\mathrm{min}}$. Since $T^{\mathrm{min}}$ decreases with decreasing $\gamma_D$, a low value for the Dynes parameter is advantageous.

\section{Transition rates in superconducting qubits}\label{sec:qubits}
In this section, we apply the general theory for the transition rates to two types of qubits: the weakly anharmonic transmon qubit and the more anharmonic capacitively shunted flux qubit. The weak anharmonicity assumption enables us to arrive at an approximate formula for the matrix elements, see Eq.~(\ref{eqn:Mtr}) (for a generic qubit, the matrix elements can always be calculated numerically, as discussed in Sec.~\ref{sec:circuitcoupling}). We aim at finding parameters which enable the use of the QCR for fast reset when in the on-state, although not causing unwanted relaxation or excitation in the off-state. Thus in the on-state, we desire that the QCR-induced relaxation rate dominates over the QCR-induced excitation rates and other excitation and relaxation rates, that is, $\Gamma_{10}^\mathrm{on} \gg \Gamma_{01}^\mathrm{on},\, 1/T_1^b$, where $\Gamma_{10/01}^\mathrm{on}$ is the bias-voltage-dependent relaxation/excitation rate of the qubit due to the QCR at the on-state and $T_1^b$ is the bare qubit relaxation time in the absence of the QCR. In the off-state, we require that the QCR has a negligible effect on the qubit, that is, $\Gamma_{10}^\mathrm{off},\,\Gamma_{01}^\mathrm{off} \ll 1/T_1^b$. As noted in Sec.~\ref{sec:circuitcoupling}, the theory developed in Ref.~\onlinecite{Silveri_2017} is applicable if, at least in the off-state, $\Gamma_{mm} \gg \Gamma_{10},\,\Gamma_{01}$, where $\Gamma_{mm}$ is the elastic tunneling rate in the QCR, not accompanied by a qubit transition.

Before going into the detailed discussion on the qubit reset, let us consider the definition of the effective temperature in Eq.~\eqref{eqn:T_T}. It follows that the finite effective temperature $T$ can limit the reset fidelity $\mathcal{F}_{r}$, 
with which the qubit 
can be prepared in the ground state. 
If the qubit relaxation rate is dominated by the QCR-induced transitions in the on-state, we have approximately $\mathcal{F}_{r} \le 1-\Gamma_{01}^\mathrm{on}/\Gamma_{10}^\mathrm{on}$. Since $\Gamma_{01}^\mathrm{on}/\Gamma_{10}^\mathrm{on}\approx e^{-2\hbar\omega_{10}/(k_B T_{N})}$ for $T_{N} > T_{N}^\mathrm{co}$, the QCR decreases exponentially the infidelity in comparison to simply waiting for equilibration at temperature $T_{N}$. For $T_{N} < T_{N}^\mathrm{co}$, we have $\Gamma_{01}^\mathrm{on}/\Gamma_{10}^\mathrm{on}\approx\gamma_D \Delta/(2\hbar\omega_{10})$, so that the reset fidelity is higher for smaller $\gamma_D$. Interestingly, even though $T^\mathrm{min}(0)$ increases with $\omega_{10}$, so does the reset fidelity.

To compare different scenarios, in addition to the reset fidelity $\mathcal{F}_r$, we report the reset time $T_{10\%}$ defined as the time in which the qubit population decreases to $10\%$ of the initial population. The reset time $T_{10\%}$ and the transition rate $\Gamma_{10}$ are related as $T_{10\%}=\ln(10)/\Gamma_{10}$. We note that the time needed to reach the maximum reset fidelity is generally somewhat longer than $T_{10\%}$, by a factor of 3 (5) at high (low) temperature $T_N$. The key results for the qubit reset are collected in Table~\ref{tab:key_results}.

\renewcommand{\arraystretch}{1.25}
\begin{table}[tb]

    \centering
        \begin{tabular}{c |c  c  |c    c|}
        & \multicolumn{2}{c|}{$T_{N}=$~\SI{10}{\milli\kelvin}} & \multicolumn{2}{c|}{$T_{N}=$~\SI{100}{\milli\kelvin}}\\
        & $1-\mathcal{F}_{r}$  & $T_{10\%}$  &  $1-\mathcal{F}_{r}$  & $T_{10\%}$\\
            \cline{1-5}
            Transmon &\num{4.3e-5} & \SI{1.5}{\nano\second} &\num{8.2e-3} & \SI{6.0}{\nano\second}  \\
                       \vtop{\hbox{\strut C-shunted}\hbox{\strut flux qubit}}  & \multirow{2}{*}{\num{5.4e-5} } & \multirow{2}{*}{ \SI{5.5}{\nano\second} } & \multirow{2}{*}{  \num{2.0e-2}} & \multirow{2}{*}{ \SI{19}{\nano\second}}
        \end{tabular}
        \caption{Key results for qubit reset: the reset infidelity $1-\mathcal{F}_{r}$ and the reset time $T_{10\%}$. The results are given at two values for the normal-metal electron temperature $T_{N}$ corresponding to the low-temperature and thermal activation regions separated by the cross-over temperature $T_N^{\rm co}=$~\SI{48}{\milli\kelvin} (transmon) and~\SI{40}{\milli\kelvin} (C-shunted flux qubit). The other  parameters are given in Table~\ref{tab:parameters}.}
        \label{tab:key_results}
    \end{table}

\subsection{Transmon}
\label{sec:Transmon}

In its simplest realization, a transmon [Fig.~\ref{fig:01}(b)] consists of a single Josephson junction with Josephson energy~$E_J$ shunted by a capacitor~$C_{\rm tr}$ to ground. Thus in the corresponding quantum circuit of Fig.~\ref{fig:QCR_circuit_gen}, there is only a single node. Due to the capacitive coupling to the QCR, the charging energy  $E_C$ of the transmon becomes slightly renormalized $E_C = e^2/\{2\left[C_{\rm tr} + C_c C_\Sigma/(C_c+C_\Sigma)\right]\}$, cf.~Appendix~\ref{app:inverseC}. Note that since typically $C_{\rm tr}$ is of the order of a few to several tens of fF, in designing the combined QCR-transmon system, it may be important to account for the coupling to the QCR, since in previous experiments $C_c \approx$~\SI{800}{\femto\farad} and $C_\Sigma \approx$~\SI{10}{\femto\farad}~\cite{Tan_2016, Masuda_2018, Silveri_2019, Sevriuk_2019}. With this notation, the Hamiltonian for the combined QCR-transmon system reads
\begin{equation}
\label{eqn:H_transmon}
\hat H_0 = \frac{1}{2C_N}\hat Q_N^2 + 4E_C (\hat n-\hat n_q)^2 - E_J \cos\hat \varphi,
\end{equation}
where $\hat n=\hat Q_0/(2e)$ and $\hat \varphi = 2\pi \hat \phi_0/\Phi_0$ are the transmon charge and flux normalized by $2e$ and the flux quantum $\Phi_0=h/(2e)$, respectively, and $\hat n_q = -\alpha \hat Q_N/(2e)$ denotes the normalized charge offset by the QCR.

In the transmon, we have $E_J \gg E_C$, and hence it can be treated as a weakly anharmonic oscillator. Thus the matrix elements $M_{mm'}$ are given, to the lowest order in the anharmonicity, by the matrix element of a harmonic oscillator, see Ref.~\onlinecite{Silveri_2017}. The corrections can be calculated analytically by treating the terms in the Taylor expansion of $\cos\hat\varphi$ beyond the quadratic term as a perturbation, cf. Ref.~\onlinecite{Koch07}. Note that by utilizing this expansion we implicitly neglect any boundary condition. Therefore, we can treat $\hat\varphi$ as a non-compact variable and use Eq.~\eqref{eqn:Mmm} for the matrix elements. The small parameter for the perturbative calculation can be expressed as
\begin{equation}\label{eqn:rho}
\rho = \pi \frac{Z}{R_K} = \sqrt{\frac{E_C}{8E_J}},
\end{equation}
where $Z=\sqrt{L/C}$ is the characteristic impedance of the effective harmonic oscillator.

Since this is a standard treatment, we do not provide a detailed derivation of the results. At next-to-leading order in $\rho$, we obtain
\begin{align}
    & M_{mm'} \simeq \langle h_m | e^{-i\frac{e}{\hbar} \alpha \hat\phi_0} | h_{m'} \rangle \label{eqn:Mtr} \\ & +  \sum_{n}\left[P_{nm'}\langle h_m | e^{-i\frac{e}{\hbar} \alpha \hat\phi_0} | h_n \rangle+P_{nm} \langle h_n | e^{-i\frac{e}{\hbar} \alpha \hat\phi_0} | h_{m'} \rangle \right], \notag
\end{align}
where the matrix elements between the harmonic oscillator states $|h_m\rangle$ are given by~\cite{Silveri_2017}
\begin{align}
& \langle h_m |e^{-i\frac{e}{\hbar} \alpha \hat \phi_0} | h_n \rangle= \notag \\
& \qquad \quad  e^{-\frac{\alpha^2 \rho}{2}} (-i)^{|l|} (\alpha^2 \rho)^{\frac{l}{2}}\left(\frac{n!}{m!}\right)^{\textrm{sgn}(l)/2} L_{n}^{|l|}(\alpha^2 \rho), \label{eq:Mho}
\end{align}
where $l=m-n$ and $L_{n}^l(\cdot)$ are the generalized Laguerre polynomials, and the perturbation matrix corresponding to the transmon anharmonicity is
\begin{align}
P_{nm} = \frac{\rho}{48}&\bigg[\delta_{n,m+4}\sqrt{(m+1)(m+2)(m+3)(m+4)} \notag \\
&+ 2\delta_{n,m+2}\sqrt{(m+1)(m+2)}(4m+6) \notag \\
&- 2\delta_{n,m-2}\sqrt{m(m-1)}(4m-2) \notag \\
&- \delta_{n,m-4}\sqrt{m(m-1)(m-2)(m-3)}\bigg]. \label{eqn:P}
\end{align}
Here we have included only the correction due to the fourth-order term in $\cos\hat\varphi$. This is sufficient for $|m-m'|\le 4$, whereas higher orders must be kept for larger index difference.  Within our approximation, in the $P_{nm'}$ ($P_{nm}$) term on the right side of Eq.~\eqref{eqn:Mtr}, we keep only the terms with $|m-n|\le 2$ ($|m'-n|\le 2$). We do not pursue this further, and only note that although the higher-order terms do not affect the power-law scaling $|M_{mm'}|^2 \propto \rho^{|m-m'|}$, they can change the numerical prefactor in comparison to the case of a harmonic oscillator.

\begin{figure}[tb]
  \includegraphics[width=1.0\linewidth]{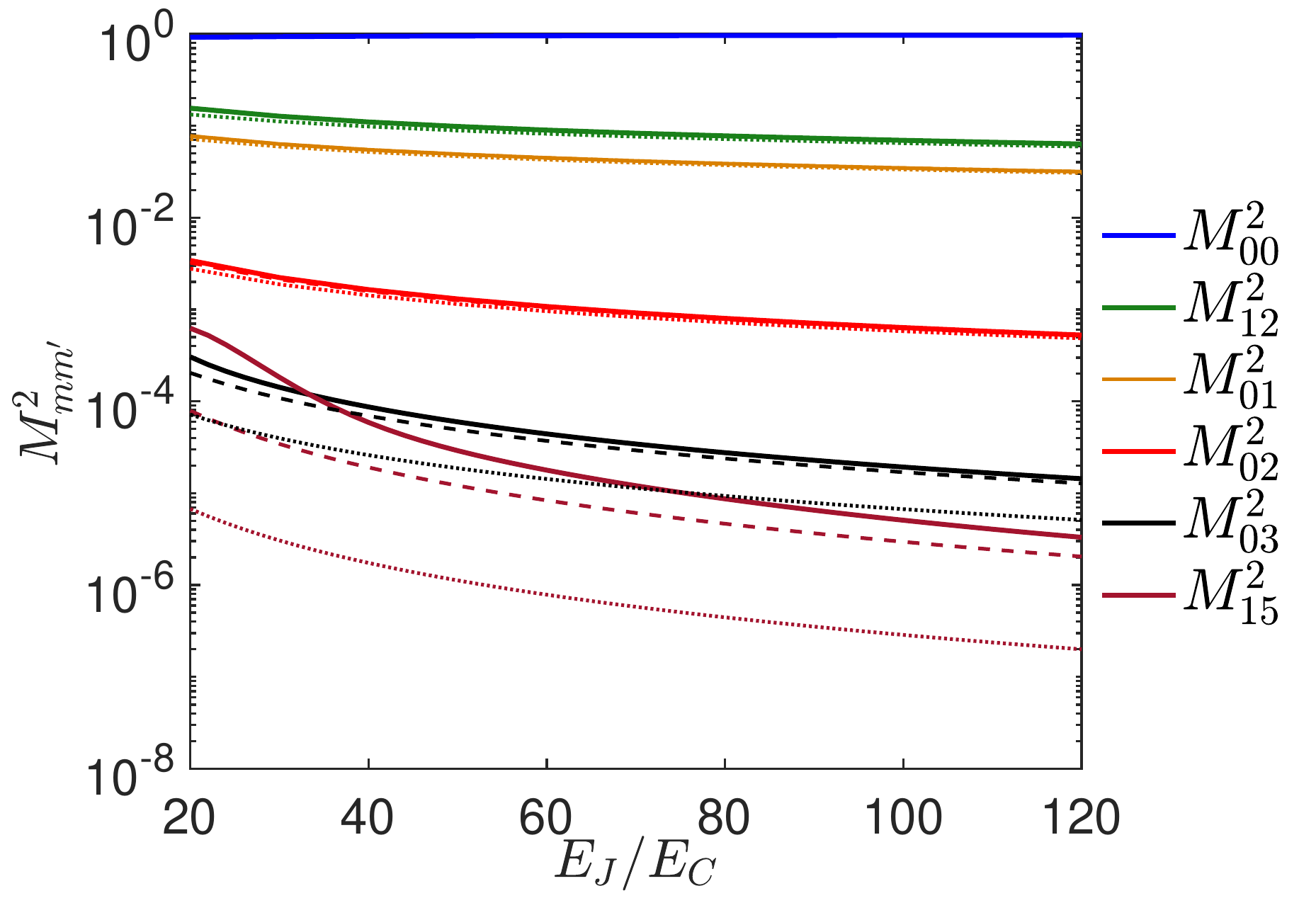}
  \caption{\label{fig:trme} Matrix elements $M_{mm'}^2$ as a function of the ratio of the Josephson energy $E_J$ to the charging energy $E_C$ for a combined QCR-transmon system. Solid lines provide the full numerical results and the dashed (dotted) lines correspond to results for an effective harmonic oscillator with (without) the anharmonic correction described in Eqs.~\eqref{eqn:Mtr}--\eqref{eqn:P}. The transition frequency $\omega_{10}/(2\pi)$ is fixed at~\SI{5}{\giga\hertz} although we change the ratio $E_J/E_C$. The other parameters are as in Table~\ref{tab:parameters}.}
\end{figure}

As shown in Fig.~\ref{fig:trme}, the calculated correction accounts for most of the deviation between the full numerical solution of the matrix elements and the exact result for the harmonic oscillator. Note that the analytical Eq.~\eqref{eqn:Mtr} is accurate down to $E_J/E_C \sim$~\num{20} for low values of $m$ and $m'$. For large $m'$, the accuracy is maintained only at large $E_J/E_C$.

Up to the correction to the matrix elements, the transition rates for the transmon are identical to those for the harmonic oscillator. Thus, we may utilize the results of Ref.~\onlinecite{Silveri_2017} for our estimates. Note, however, that the characteristic impedance for a typical transmon is larger than that of a typical coplanar waveguide resonator by roughly $0.5-1$ orders of magnitude, depending on $E_J/E_C$. Therefore, the qubit matrix element $|M_{01}|^2 \propto \rho \propto Z$ is usually greater than the corresponding matrix element for a resonator.
An important difference between a transmon and a harmonic oscillator is that only in a transmon there is charge dispersion, i.e., the energy levels as well as matrix elements depend on the initial charge offset. However, this dependence is weak by design in a transmon, and we have verified numerically that the relative change of the matrix element $M_{01}$ with initial charge offset is of order \SI{0.1}{\percent} for $E_J/E_C$ as low as \num{20}, and decreases exponentially when increasing this ratio. Because of the charge dispersion, moreover, charge fluctuations due to the QCR can lead to dephasing\----we consider this issue in Sec.~\ref{sec:dephasing}. Here, we focus on determining the useful parameter regime for operating the combined QCR-transmon system.

\subsubsection{Off-state transition rates}

Let us consider the QCR in the off-state at $eV= 0$. Here, the effective bath temperature is of the order of the normal-metal island temperature, $T \simeq T_N$, and hence for $k_B T_N \ll \hbar \omega_{10}$, which we always assume, we have by the detailed balance
\begin{equation}\label{eq:gammadetbal}
    \Gamma_{01}^\mathrm{off}/\Gamma_{10}^\mathrm{off} = e^{-\hbar \omega_{10}/(k_B T)} \ll 1 \,.
\end{equation}
Therefore, we only have to check the condition $\Gamma_{10}^\mathrm{off} \ll 1/T_1^b$. From Eq.~\eqref{eqn:Gamma}, keeping the leading term in the matrix element Eq.~(\ref{eqn:Mtr}), and using Eq.~(\ref{eq:F_0T}), we estimate (cf. Ref.~\onlinecite{Silveri_2017})
\begin{equation}
\label{gamma_off}
\Gamma_{10}^\mathrm{off} \simeq  \bar\gamma \gamma_D \frac{\Delta}{\sqrt{\Delta^2-(\hbar \omega_{10})^2}} \approx \bar\gamma \gamma_D,
\end{equation}
where $\bar \gamma$ is the asymptotic, $eV/\Delta \gg 1$, coupling strength
\begin{equation}
\label{eqn:gammaT_bar}
\bar\gamma = \frac{2}{\pi}\omega_{10} g_T \alpha^2\rho,
\end{equation}
with $g_T =R_K/R_T$ being the dimensionless conductance of the QCR normal-metal--insulator--superconductor junctions. The off-state transition rate $\Gamma_{10}^\mathrm{off}$ is finite due to the subgap conductance of the NIS junction characterized by the finite Dynes parameter.

For the typical transmon parameters in Table~\ref{tab:parameters}, we obtain $\rho = 0.05$ according to Eq.~\eqref{eqn:rho}, $g_T=0.5$, and $\bar\gamma\approx \alpha^2\times 5\times 10^{8}$~1/s. Since the relaxation time $T_1^b$ is of the order of a few hundred microseconds~\cite{Paik11, Rigetti_2012, Chang_2013} for state-of-the-art transmons, and we find an upper limit for the Dynes parameter in this case to be $\gamma_D < 10^{-5}/\alpha^2$.  For $\alpha=1$, this value is smaller than those already obtained in experiments with a QCR~\cite{Tan_2016, Masuda_2018, Silveri_2019, Hyyppa19}, but in general, values down to \num{e-7} have been demonstrated~\cite{Pekola10}. If it appears challenging to reduce the Dynes parameter to such a low value in practice, it is straightforward to reduce $\alpha$ by reducing the coupling capacitance between the qubit and the QCR in fabrication. Hence it is always possible to ensure that in the off-state, the QCR does not adversely affect the qubit.

Consequently, one should try to maximize the on/off ratio of the QCR. Note that if the Dynes broadening is caused by the weak proximity effect at the normal-metal--insulator--superconductor junction, we have $\gamma_D \propto g_T$~\cite{Hosseinkhani_2018}, and hence we can reduce $\gamma_D$ by fabricating more opaque tunnel barriers. As discussed below, a lower Dynes parameter yields a higher on/off ratio in general.

\subsubsection{On-state reset fidelity and reset time}

\begin{figure}[tb]
    \centering
    \includegraphics[width=.95\linewidth]{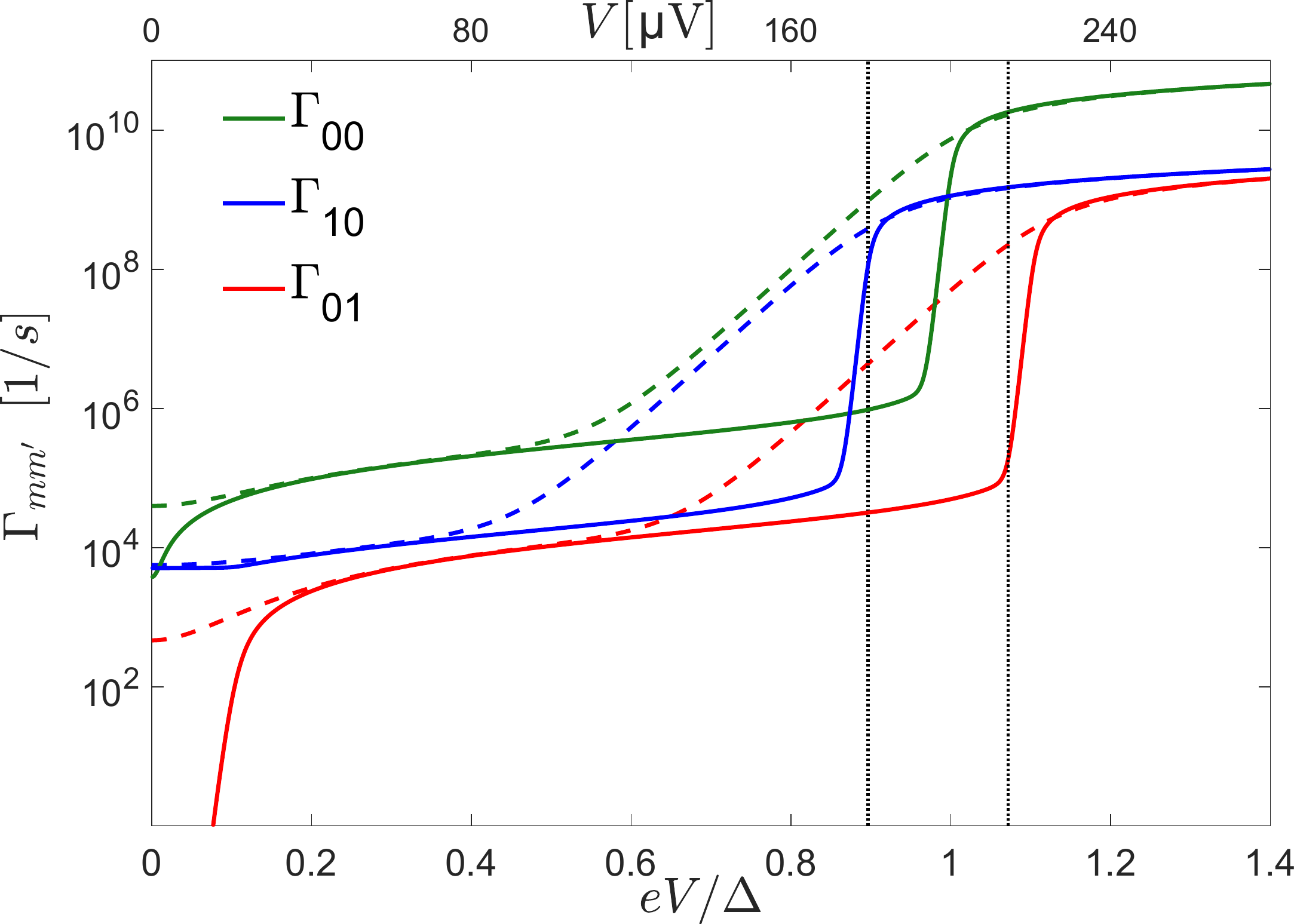}
    \caption{\label{fig:04} Transition rates $\Gamma_{mm'}$ as a function of the bias voltage $V$ in a combined QCR and transmon system for the normal-metal electron temperature $T_N=$~\SI{100}{\milli\kelvin}~(dashed lines) and $T_N=$~\SI{10}{\milli\kelvin}~(solid lines). The vertical black dotted lines denote, from left to right, the on-voltage in the thermal activation regime $eV=\Delta-\hbar \omega_{10}$ and  the on-voltage in the low-temperature regime $eV=eV_\mathrm{max}$ of Eq.~(\ref{eq:eVmax}) at $T_N=$~\SI{10}{\milli\kelvin}.
    The transmon and QCR parameters are given in Table~\ref{tab:parameters}.
    }
\end{figure}

We treat the two regimes of $T_{N}$ above or below $T_{N}^\text{co}$ separately, as already discussed in Sec.~\ref{sec:operation}. For $T_{N}>T_{N}^{\rm co}$, we consider the operation point $eV = \Delta-\hbar \omega_{10}$, since at this point we have the largest coupling strength compatible with $T\approx T_{N}/2$. At this value of bias, we find (see Appendix~\ref{app:onoff})
\begin{equation}\label{eq:GonhighT}
    \Gamma_{10}^\text{on} \approx  0.38\times \bar{\gamma} \frac{\sqrt{k_B T_{N}\Delta}}{\hbar \omega_{10}}\, ,
\end{equation}
and using the parameters from Table~\ref{tab:parameters}, we estimate the transition rate $\Gamma_{10}^\mathrm{on} =  0.4\textrm{ }1/\textrm{ns}$ for $T_N=100$~mK. 
Figure~\ref{fig:04} shows an example of transition rates as a function of the bias voltage at a temperature $T_{N}>T_{N}^{\rm co}$. The rates are small compared to $1/T_1^b$ at low bias, and increase exponentially as the bias $eV$ increases towards the superconducting gap $\Delta$.

We estimate the reset fidelity $\mathcal{F}_{r}$ by considering a qubit initially in the excited state, and calculating the probability that it is found in the ground state after a certain reset time $t_r$ has passed. We find approximately the reset infidelity $$1-\mathcal{F}_r \approx e^{-\Gamma_{10}^\mathrm{on} t_r}+\Gamma_\uparrow/\Gamma_{10}^\mathrm{on},$$ where $\Gamma_\uparrow$ is the total qubit excitation rate due to both the QCR and other sources.
The last term is the ratio of the excitation to the relaxation rate, for which there are two possible sources. The first source is the QCR itself, for which we have by the detailed balance $\Gamma_{01}^\mathrm{on}/\Gamma_{10}^\mathrm{on}= e^{-2\hbar \omega_{10}/(k_B T_N)}\approx$~\num{8.2e-3} (using the parameters in Table~\ref{tab:parameters} for $T_N=100$~mK). 
The second source is due to other processes, and their excitation rate can be estimated as $P_e/T_1^b$, where $P_e$ is the steady-state probability for the qubit to be in the excited state, typically ranging from a fraction of a percent~\cite{Jin_2015} to about \SI{10}{\percent}~\cite{Geerlings_2013a}.
Taking for example $P_e=$~1\%, we find $P_e/(\Gamma_{10}^\mathrm{on}T_1^b) \lesssim$~\num{e-6}. Thus in this temperature regime, the fidelity of initialization is limited by the rates of the QCR.

The reset time for $T_N=100$~mK is $T_{10 \%}=6$~ns which is in stark contrast with the a few hundred microseconds for the natural reset of transmons.
The obtained reset time is comparable to single- and two-qubit gate times and is potentially shorter than that of the alternative proposals for qubit reset~\cite{Reed_2010,Riste_2012,Johnson_2012,Campagne_2013,Salathe_2018,Valenzuela_2006,Grajcar_2008,Geerlings_2013a,McClure_2016,Magnard_2018,Egger_2018}. Our conclusions hold even if the qubit has somewhat lower or higher frequency, and if we allow for qubit thermalization~\cite{Jin_2015} to increase $P_e$, while keeping $T_1^b$ at least in the microsecond regime. However, a low excited state population is desirable for quantum computation, and this motivates us to pursue lower $T_{N}$.

We expect the QCR to be typically operated at the low-temperature regime such that $T_{N}<T_{N}^{\rm co}$. 
In this case, the operation point at the on-state is $eV \simeq eV_\mathrm{max} > \Delta$ of Eq.~(\ref{eq:eVmax}).
We use Eq.~(\ref{eq:F_0T}) and Eq.~(\ref{eqn:Gamma}) to estimate the transition rate as
\begin{equation}\label{eq:GammaonlowT}
    \Gamma_{10}^\mathrm{on} \approx \frac{\bar{\gamma}}{2\hbar \omega_{10}} \sqrt{\left(eV_\text{max}+\hbar \omega_{10}\right)^2-\Delta^2}.
\end{equation}
For the parameters in Table~\ref{tab:parameters} and $T_N=10$~mK, we have $\Gamma_{10}^\mathrm{on} = 1.5/\textrm{ns}$; the corresponding reset time $T_{10\%} = 1.5$~ns, 
which is again several orders of magnitude shorter than $T_1^b$. We also obtain $T^{\rm min}(0) \approx$~\SI{24}{\milli\kelvin} from Eq.~(\ref{eqn:T^min_T}) and by the detailed balance $\Gamma_{01}^\mathrm{on}/\Gamma_{10}^\mathrm{on} = e^{-\hbar \omega_{10}/(k_B T^{\rm min})} =$~\num{4.3e-5}, the excitation rate due to the QCR  dominates over those of other processes. Therefore, the reset infidelity  $$1-\mathcal{F}_r = \Gamma_{01}^\mathrm{on}/\Gamma_{10}^\mathrm{on},$$ is adequately small.
Thus, by lowering the electron temperature $T_{N}$ the reset fidelity is improved, since $1-\mathcal{F}_r$ is reduced by one to two orders of magnitude. Simultaneously, the transition rate and the reset time are enhanced by roughly a factor of four. 

The improvements obtained by lowering the normal-metal electron temperature $T_N$ as discussed above are visible in Fig.~\ref{fig:04}, where we compare the transition rates at $T_N=100$~mK to those at $10$~mK. Indeed, the relaxation rate $\Gamma_{10}$ at the low-temperature regime on-voltage $V=V_\mathrm{max}$ is about \num{14} times that for the on-voltage $eV=\Delta-\hbar \omega_{10}$ at the thermal activation regime, where as the excitation rate $\Gamma_{01}$ is increased by about a factor of five. Thus the reset is faster and has higher fidelity at low $T_N$ compared with high temperature.

\subsubsection{On/off ratio}
At $T_N\ll T_N^\textrm{co}$ and for $\hbar\omega_{10}\ll \Delta$, we may use Eqs.~\eqref{eq:GammaonlowT} and~\eqref{gamma_off} to express the QCR on/off ratio as
\begin{equation}\label{eq:onoff}
    R_\textrm{on/off}=\Gamma_{10}^\textrm{on}/\Gamma_{10}^\textrm{off}\approx \sqrt{\frac{\Delta}{\hbar\omega_{10}\gamma_D^2}}.
\end{equation}
Thus, if one chooses the parameters of the QCR such that in the off state, the decay introduced by the QCR is somewhat below that of the other sources, one obtains that $\Gamma_{10}^\textrm{on}\simeq 1/(\gamma_{D} T_1^b)$. Or equivalently, the qubit $T_1$ is multiplied by a factor of roughly $\gamma_D$ in the on-state. This analysis highlights the importance of reaching low Dynes parameters. 

\subsection{Capacitively shunted flux qubit}
\label{sec:CSFQ}

We consider a flux qubit consisting of three Josephson junctions [Fig.~\ref{fig:01}(b)], two of which with Josephson energy $E_J$ and junction capacitance $C_J$ and the remaining junction with a Josephson energy $\alpha_\textrm{fl}E_J$ and capacitance $\alpha_\textrm{fl}C_J$, where $\alpha_\mathrm{fl}$ is a constant. Referring to the quantum circuit of Fig.~\ref{fig:QCR_circuit_gen}, there are two nodes and the QCR is capacitively coupled to one of these. The QCR-renormalized inverse of the matrix $\boldsymbol{C}$ can be obtained from Eq.~\eqref{eqn:Cinv}, where $\widetilde{C}_0=C_J(1+2\alpha_\mathrm{fl})/(1+\alpha_\mathrm{fl})+C_c C_\Sigma/(C_c+C_\Sigma)$. Note that in a regular flux qubit the junction capacitance $C_J$ can be typically of the order of a few to a few tens of~fF~\cite{Stern14,Yan16}. Thus, in the point of view of the node charging energy, the renormalization of the order of~\SI{10}{\femto\farad} by the QCR is significant. The reduced charging energy is reminiscent to that of the so-called capacitively shunted flux qubit~\cite{Yan16}, where the small junction is shunted with an additional capacitance $C_{\rm sh}$ in the range \SIrange{10}{50}{\femto\farad}. 
Conveniently, we consider below only a capacitively shunted flux qubit with the shunt capacitance $C_\mathrm{sh}=\zeta C_J$.

The Hamiltonian of the combined system of a capacitively shunted flux qubit~\cite{Yan16} and the QCR is given by
\begin{align}
  \hat H_0  = &\frac{\hat Q_N^2}{2C_N} +
 \frac{2e^2}{C'_J}\bigg\{(1+\alpha_\mathrm{fl}+\zeta)\left(\hat n_1+\frac{\hat n_q}{2}\right)^2\notag \\& + 2(\alpha_\mathrm{fl}+\zeta)\left(\hat n_1+\frac{\hat n_q}{2}\right) \hat n_2 \notag \\ & +\left[ 1+\alpha_\mathrm{fl}+\zeta+\frac{C_c C_\Sigma}{C_J (C_c+C_\Sigma)} \right] \hat n_2^2 \bigg\} +\hat U_{\rm fl}, \label{eqn:H_flux}
\end{align}
where the potential energy is given by
\begin{align}
   \hat U_{\rm fl}=E_{J} \big[2 +\alpha_\mathrm{fl} &-\cos\hat \varphi_1-\cos\hat \varphi_2 \notag \\ & -\alpha_\mathrm{fl}\cos(2 \pi f_e +\hat \varphi_1-\hat \varphi_2) \big],
\end{align}
and the capacitance renormalization by the QCR appears in the effective junction capacitance
\begin{equation}\label{eqn:CJp}
  C'_J =C_J(1+2\alpha_\mathrm{fl}+2\zeta)+\frac{C_c C_\Sigma}{C_c+C_\Sigma}(1+\alpha_\mathrm{fl}+\zeta).
\end{equation}
In capacitively shunted flux qubits, the junction capacitance $C_J$ is typically of the order of several tens of~fF, which is larger than $C_c C_{\Sigma}/(C_c + C_{\Sigma})$. Since $1+\alpha_\mathrm{fl}+\zeta > 1$ and $C_c C_{\Sigma}/[(C_c + C_{\Sigma})C_J] \lesssim 1$, we may neglect the last term in the square brackets in Eq.~\eqref{eqn:H_flux}, especially if $\zeta \gg 1$. We use this option below to simplify the problem. However, we take exactly into account the renormalization due to the last term in Eq.~\eqref{eqn:CJp}.

\begin{figure}[tb]
  \centering
  \includegraphics[width=1.0\linewidth]{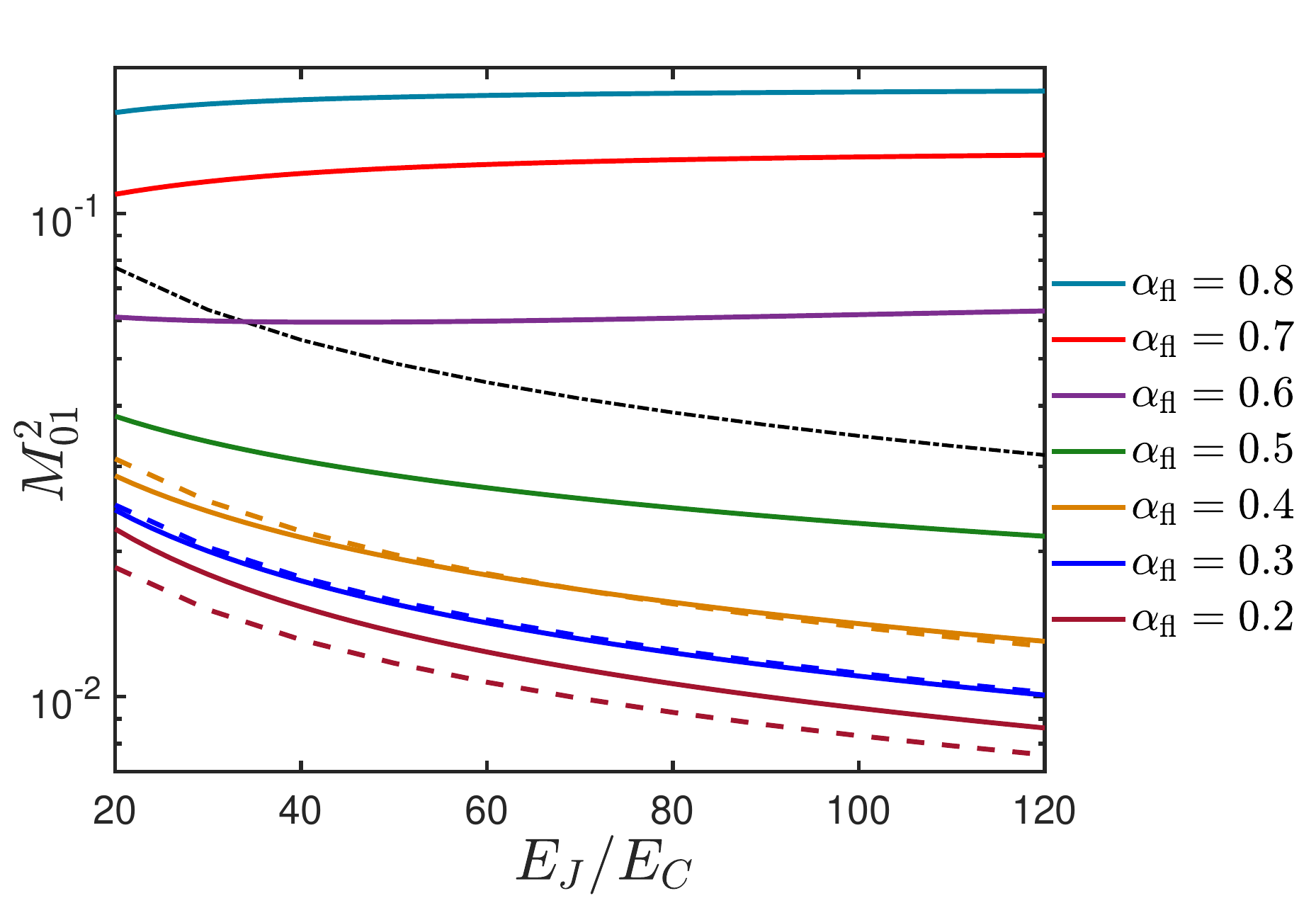}
  \caption{\label{fig:CSFQ_me_EJdEC} Matrix element $M^2_{01}$ of a capacitively shunted flux qubit (solid lines) as a function of the ratio $E_J/E_C$ between the charging energy and the Josephson energy. We fix the parameters 
  according to Table~\ref{tab:parameters} except that we
  vary $\alpha_{\rm fl}$ as indicated. The dashed lines are obtained by utilizing the 1D model of a capacitively shunted flux qubit in Eq.~\eqref{eqn:flux1D} within the harmonic approximation in Eq.~\eqref{eq:Mho} and $\rho$ given by Eq.~\eqref{eqn:effrho} for $\alpha_\mathrm{fl}=0.2,~0.3,~0.4$, respectively from bottom to top.
  For higher values of $\alpha_\mathrm{fl}$, the qubit potential generally has a double-well shape, such that a comparison to a transmon or a harmonic oscillator is not productive. Indeed, the matrix elements can even increase rather than decrease with increasing~$E_J/E_C$. The black dash-dotted line shows the matrix element of a corresponding transmon calculated in the harmonic approximation, Eq.~\eqref{eq:Mho}, with $\rho$ given by Eq.~\eqref{eqn:rho}.
}
\end{figure}

Next, we compare the matrix elements $|M_{01}|^2$ of capacitively shunted flux qubits and transmons in Fig.~\ref{fig:CSFQ_me_EJdEC}. For capacitively shunted flux qubits, we calculate the matrix elements $M_{mm'} = \langle m'_{q+1} \ket{m_q}$ numerically, as explained at the end of Sec.~\ref{sec:circuitcoupling}. For $\alpha_\mathrm{fl}<0.5$, a typical regime for a capacitively shunted flux qubit, the matrix element $|M_{01}|^2$ decreases as the ratio $E_J/E_C$ increases similarly as for a transmon but the magnitude is smaller. This can be explained using an effective one-dimensional (1D) model. As discussed above, we neglect the last term in the square brackets in Eq.~\eqref{eqn:H_flux}. We also employ a coordinate rotation $\hat \varphi_p=(\hat \varphi_1+\hat \varphi_2)/2$ and $\hat \varphi_m=(\hat \varphi_1-\hat \varphi_2)/2$ to diagonalize the capacitance matrix. If $\zeta \gg 1$, the oscillatory modes in the $\hat \varphi_p$ direction turn out to have high energies. We neglect such modes~\cite{Yan16}, since we are interested in the low-energy eigenstates, the properties of which are approximately described by the Hamiltonian
\begin{align}
     \hat H_{\rm fl, 1D}& =\frac{1}{2}E_{C,m}\left(\hat n_m+\frac{\hat n_q}{2}\right)^2 \label{eqn:flux1D}\\
       + & E_J \bigg\{ -2\cos\hat \varphi_m+\alpha_\mathrm{fl}\cos \bigg[ 2\pi\left(f_e-\frac{1}{2}\right)+2\hat \varphi_m \bigg] \bigg\},\notag
\end{align}
where $E_{C,m}=2e^2/C_J'$ is the charging energy of the oscillatory modes in the $\varphi_m$ direction.
For $f_e=$~\num{0.5} and $\alpha_\mathrm{fl}<$~\num{0.5}, we may expand the cosine terms around $\varphi_m=0$ and keep only the quadratic terms to obtain
\begin{equation}
\label{eqn:flux1Dexpand}
\hat{\tilde{H}}_{\rm fl, 1D}=\frac{1}{2}E_{\rm C,m}\left(\hat n_m+\frac{\hat n_q}{2}\right)^2+ E_J (1-2 \alpha_\mathrm{fl})  \hat \varphi_m^2.
\end{equation}
Comparing Eq.~\eqref{eqn:flux1Dexpand} with Eqs.~\eqref{eqn:H_transmon}--\eqref{eqn:rho}, we find the parameter $\rho$ of a capacitively shunted flux qubit to be
\begin{align}
\label{eqn:effrho}
& \rho=\sqrt{\frac{E_C}{8E_J}} \Bigg\{ \frac12 \frac{1}{\sqrt{1-2\alpha_\mathrm{fl}}} \\ & \times \Bigg[1 + 2\alpha_\mathrm{fl}+2\zeta + \frac{C_c C_\Sigma}{C_J(C_c+C_\Sigma)}(1+\alpha_\mathrm{fl}+\zeta) \Bigg]^{-\frac12} \Bigg\}, \nonumber
\end{align}
where $E_C=e^2/(2C_J)$.
Therefore the matrix element $|M_{01}|^2$, which is proportional to $\rho$, decreases with increasing $E_J/E_C$, similarly to the transmon, but the magnitude is smaller due to the coefficient in the curly brackets, see Fig.~\ref{fig:CSFQ_me_EJdEC}.
The matrix elements of capacitively shunted flux qubits decrease with increasing $\zeta$. This is not surprising since increasing $\zeta$ effectively decreases the charging energy.

Note the good agreement between our numerical results and the harmonic approximation despite the stronger relative anharmonicity for flux qubits with $\alpha_\mathrm{fl} >  0.2$ than for transmons: Whereas for the latter the relative anharmonicity is given by $-\rho$ of Eq.~(\ref{eqn:rho}), for the former it is $\rho (8\alpha_\mathrm{fl}-1)/(1-2\alpha_\mathrm{fl})$ with $\rho$ of Eq.~(\ref{eqn:effrho}), as can be checked by considering the fourth order term in the expansion of Eq.~(\ref{eqn:flux1D}) around $\varphi_m =0$ for $f_e =0.5$~\cite{Yan16}.

\subsubsection{Off-state transition rates}

Here we discuss the upper bound for the Dynes parameter for the QCR to work in practice. The parameters of the capacitively shunted flux qubit are given in Table~\ref{tab:parameters} according to Ref.~\onlinecite{Yan16} and our corresponding results are summarized in Table~\ref{tab:key_results}. 
The small parameter $\rho$ obtained from Eq.~\eqref{eqn:effrho} equals \num{0.016}, which is smaller than what we have for a transmon, owing to a larger $E_J/E_C$ and to the renormalization in capacitively shunted flux qubits. As in Fig.~\ref{fig:CSFQ_me_EJdEC}, the 1D model of Eq.~\eqref{eqn:flux1Dexpand} 
agrees well with the full 2D model of Eq.~\eqref{eqn:H_flux} around $\alpha_\mathrm{fl}=0.4$. Since in this regime, the 1D model can be accurately approximated by that of a harmonic oscillator, we use the theory developed in Ref.~\onlinecite{Silveri_2017} for a harmonic system to calculate the coupling strength as we did for a transmon. Specifically, we can use Eq.~(\ref{eqn:gammaT_bar}) to obtain the asymptotic coupling strength
$\bar \gamma = 0.13\times 10^9$~1/s, corresponding to $1/\bar \gamma \approx 8$~ns. For a capacitively shunted flux qubit, the typical relaxation time $T_1^b$ is tens of microseconds~\cite{Yan16}, a factor of ten shorter than the best transmons. In turn this relaxes the upper limit for the Dynes parameter ensuring that $\Gamma_{10}^{\rm off}<1/T_1^b$ and yielding now $\gamma_D<10^{-4}$, which is larger by one order of magnitude than that for transmons. However, to ensure high reset fidelity, a low value of $\gamma_D$ is needed. For our estimates below, we use $\gamma_D = 10^{-5}$, as in Sec.~\ref{sec:operation}.

\subsubsection{On-state reset fidelity and reset time}

Let us estimate the reset time and reset fidelity in the on-state for the parameters of Table~\ref{tab:parameters}. In the thermal activation regime of Eq.~(\ref{eq:GonhighT})  where the normal-metal electron temperature $T_{N}=$~\SI{100}~mK~$>T_{N}^{\rm co}$, we have the transition rate $\Gamma_{10}^\mathrm{on}  = 0.12\,1/\textrm{ns}$ yielding the reset time $T_{10\%}\approx 19$~ns. Similarly to a transmon, the reset fidelity is limited by the QCR induced excitations $1-F_{r}= \Gamma_{01}^\mathrm{on}/\Gamma_{10}^\mathrm{on} \approx 0.02$. 

In the low-temperature regime, where $T_{N}=10\textrm{ mK}<T_{N}^{\rm co}$, we use Eq.~\eqref{eq:GammaonlowT} to obtain $\Gamma_{10}^\mathrm{on} = 0.42\textrm{ }1/\textrm{ns}$ and $T_{10\%}= $~\SI{5.5}{\nano\second}$\,\ll T_1^b$. The minimum effective temperature at zero temperature $T^{\rm min}(0)$ is~\SI{20}{\milli\kelvin} for $\omega_{10}/2 \pi=$~\SI{4.1}{\giga\hertz} resulting in the limit to the reset infidelity $1-\mathcal{F}_r = e^{-\hbar \omega_{10}/(k_B T^{\rm min})} \approx$ \num{5.4e-5}.  In summary, we find that our estimates for reset fidelities are comparable for both capacitively shunted flux qubits and transmons. However, for capacitively shunted flux qubits due to the reduced matrix elements, the relaxation rates are lower and therefore reset times longer.

\subsection{Leakage and anharmonicity}

\begin{figure}[tb]
  \centering
  \includegraphics[width=1.0\linewidth]{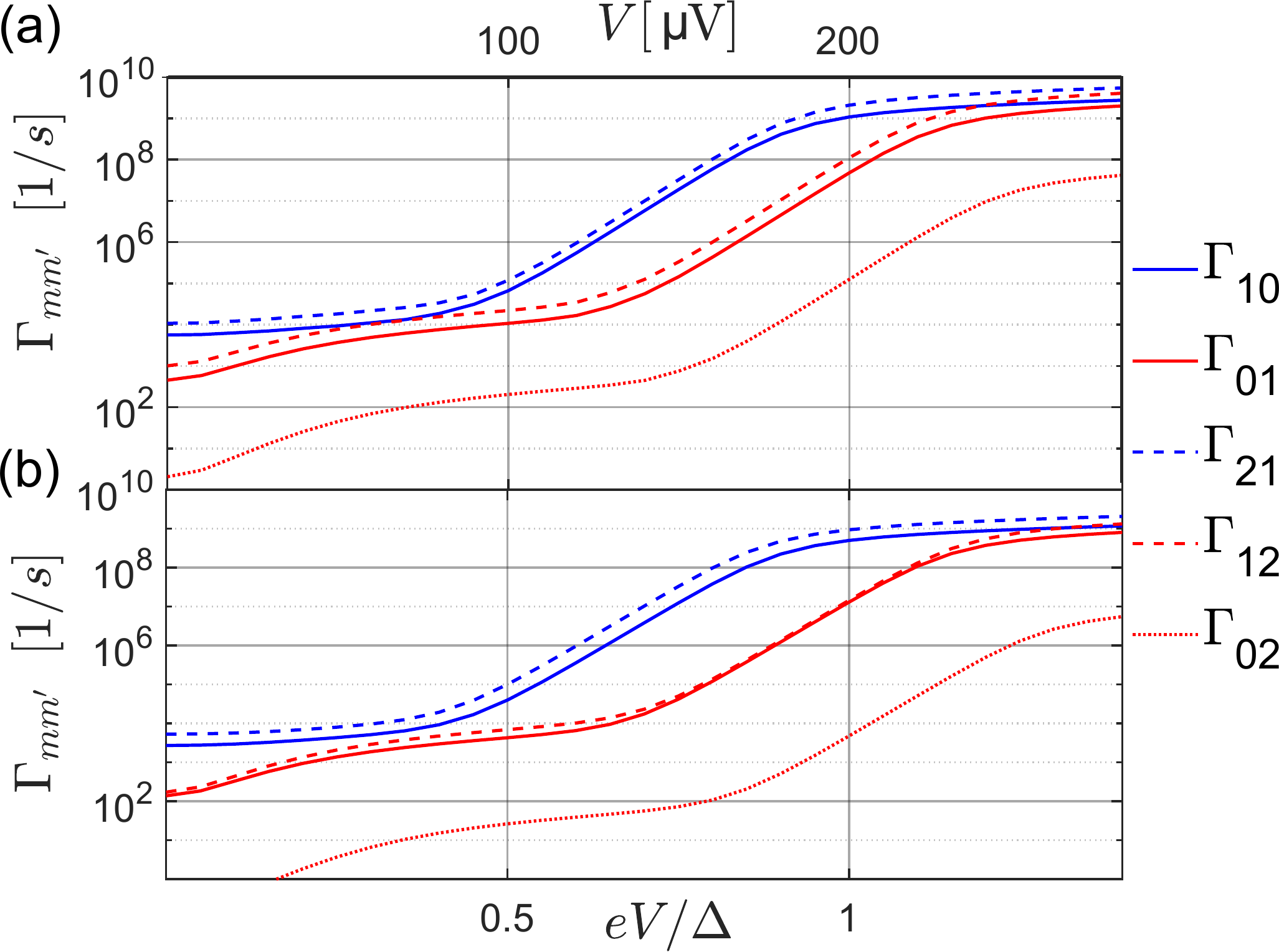}
  \caption{\label{fig:06} Transition rates $\Gamma_{mm'}(V)$ as a function of the bias voltage $V$ in (a) a QCR-transmon system and in (b) a QCR--capacitively-shunted-flux-qubit system. For both qubit types, the ratio $E_{J}/E_{C}=$~\num{50}. 
  The transition frequency and anharmonicity of the transmon (capacitively shunted flux qubit) are $\omega_{10}/2 \pi=$~\num{5}~GHz (\num{6}~GHz) and \num{-0.303}~GHz (\num{0.912}~GHz), respectively. For the QCR, we use 
  the electron temperature $T_N=$~\SI{100}{\milli\kelvin}. The other parameters are as in Table~\ref{tab:parameters}.}
\end{figure}

Above, we focused on reset times and fidelity. The QCR can also cause or mitigate unwanted transitions in the quantum circuit, for example,  to the second excited state. Here, we show that such leakage transitions have a negligible effect on the reset fidelity. In discussing the $0\to 1$ transition rates, we have already shown that $\Gamma_{01} \ll \Gamma_{10}$, both for off- and on-states. Thus the most important unwanted transitions are $0\to 2$ and $1\to 2$. Treating both qubit types within the harmonic approximation, we have $|M_{12}|^2/|M_{01}|^2 \approx 2$ and $\omega_{21} \approx \omega_{10}$. Therefore we estimate $\Gamma_{12}/\Gamma_{01} \approx \Gamma_{21}/\Gamma_{10} \approx 2$, independent of the QCR bias voltage. This implies that $\Gamma_{12} \ll \Gamma_{10}$, and hence leakage from the excited state is suppressed. Taking into account the anharmonicity does not qualitatively change this result: a small negative anharmonicity such as that in a transmon is equivalent to applying a slightly higher voltage for the $1 \to 2$ transition, and hence somewhat increasing the transition rate with respect to our estimate. We have verified that this increase in the rate $\Gamma_{12}$ does not affect our conclusion as exemplified in Fig.~\ref{fig:06}. On the contrary, for positive anharmonicity the rate $\Gamma_{12}$ is slightly suppressed. Positive anharmonicity is found in a capacitively shunted flux qubit with $\alpha_\mathrm{fl}> 1/8$, see the discussion in the paragraph after that containing Eq.~(\ref{eqn:effrho}).

For the $0\to2$ transition, we note that in the harmonic approximation $|M_{02}|^2/|M_{01}|^2 \propto \rho \ll 1$, implying a suppression of the rate  $\Gamma_{02}$ compared to $\Gamma_{01}$. Moreover, the higher energy difference of the transition $\omega_{20} \approx 2\omega_{10}$ further reduces the transition rates. Therefore we can safely neglect the $0\to2$ transition. Similar arguments can be used to show that transitions to even higher levels are irrelevant in our regime of interest. Finally, restricting our attention to the three lowest levels and including only the most important, nearest-level transitions, we find that the steady-state occupation probability of the ground states is approximately $1-\Gamma_{01}/\Gamma_{10}(1+\Gamma_{12}/\Gamma_{21}$); since as we discussed $\Gamma_{12}/\Gamma_{21} \ll 1$, we find a negligible suppression of the reset fidelity due to leakage.

In the case of quantum error correction however, the significance of the leakage has to be reconsidered. Even a tiny relative leakage error of $10^{-6}$ or smaller may become dominating if its effect to the logical qubit states is not corrected for. We leave this topic for future research.

\section{Dephasing by temporal charge fluctuations}  \label{sec:dephasing}

In Secs.~\ref{sec:operation} and~\ref{sec:qubits}, we have concentrated on how the quantum-circuit refrigerator induces transitions in the coupled quantum circuit. The quantum-circuit refrigerator has also an effect on the eigenenergies of the quantum circuit. To this end, we can identify two different contributions: a deterministic shift and temporal fluctuations. The deterministic shift can be understood as a combination of Lamb and ac Stark shifts due to coupling to an electromagnetic environment~\cite{Silveri_2019}, considered in detail for general anharmonic systems in Sec.~\ref{sec:Lamb}.  In this section, we focus on the temporal fluctuations caused by the fluctuating charge on the normal-metal island of the QCR. We are especially interested in minimizing the dephasing of the quantum circuit induced by these temporal fluctuations.

Considering the Hamiltonian~\eqref{eqn:HC} we define the eigenenergies of the quantum circuit $E_{m}(n_q)$ as a function of the normalized charge of the normal-metal island, $n_q=-\alpha Q_N/(2e)$. We define the charge dispersion of the $m$th eigenstate as $\tilde\varepsilon_m=\max_{n_g}E_m(n_g)-\min_{n_g}E_m(n_g)$. 
We are interested in superconducting quantum devices where the Josephson relation renders the energies periodic in units of the charge of a Cooper pair and thus the charge dispersion is a well defined quantity. Note that the charge $Q_N$ of the normal-metal island fluctuates in a discrete manner in units of $e$ due to electron tunneling through the normal-metal--insulator--superconductor junction, implying discrete fluctuations by $\pm \alpha/2 $ in $n_q$. This implies discrete temporal fluctuation of the energies $E_m(n_q)$ by an amount bounded  by the charge dispersion $\tilde\varepsilon_m$. This kind of process is generally referred to as random telegraph noise, which is thoroughly studied due to its simplicity and physical relevance starting from early nuclear magnetic-resonance spectroscopy studies~\cite{Anderson_1954, Abragam, Li_2013}.

For simplicity, we consider telegraph noise which is characterized by the jump rate $\chi$ and the jump amplitude $\xi$ defined such that the transition energy between two eigenstates is $\hbar\omega_{nk}(t)=E_n-E_k \pm \hbar \xi$. 
Telegraph noise causes dephasing in the system with two distinct regimes denoted as the slow- and fast-jumping regimes. In the slow-jumping regime, the jump rate is much lower than the jump amplitude, $\chi\ll\xi$, and hence the induced additional dephasing rate by the telegraph noise is given by $\chi$. Thus the total dephasing rate is $\Gamma_2'=\Gamma^b_2+\chi$, where $\Gamma^b_2$ is the dephasing rate in the absence of the telegraph noise. The dephasing time shortens accordingly $T_2'=1/(\Gamma^b_2+\chi)$.

In the fast-jumping regime, the jump rate exceeds the jump amplitude, $\chi\gg \xi$, and the induced additional dephasing is reduced to $\xi^2/(2\chi)$ thanks to a phenomenon referred to as motional narrowing~\cite{Anderson_1954, Abragam, Li_2013}. The total dephasing rate is $\Gamma_2'=\Gamma^b_2+\xi^2/(2\chi)$ in the fast-jumping limit. The motional averaging can be intuitively understood through energy--time uncertainty relations~\cite{Sakurai}.

In our case, the telegraph noise is caused by all the tunneling transitions that change the charge state of the normal-metal island. Hence the jump rate is the total rate of tunneling events including both elastic and inelastic events. The rate of elastic tunneling events typically dominates over the rate of inelastic tunneling events or is at least roughly at the same scale~\cite{Silveri_2017}, see Fig.~\ref{fig:04}. Thus we characterize the jump rate by the rate of elastic tunneling events $\Gamma_{mm}$ of Eq.~\eqref{eqn:Gamma}. Assuming weak interaction $\rho\ll 1$ and a weakly anharmonic quantum circuit, the elastic rates $\Gamma_{mm}=\Gamma_{\rm el}$ become independent of the state
\begin{equation}\label{eqn:elas}
 \Gamma_{\rm el}\approx\frac{2 R_{K}}{R_{T}}\sum_{\tau=\pm 1}F(\tau eV),
\end{equation}
where we have ignored the small shift by the normal-metal charging energy $E_{N}$ for simplicity. At the on-state, we are interested in resetting the qubit where dephasing plays no role. Thus, we focus here on the dephasing properties at the off-state which covers majority of the operation times. Consequently, it is important to understand the dephasing properties of the off-state in detail. We discuss briefly the case of the on-state in Appendix~\ref{app:on_deph}.

At the off-state defined by low bias voltages $|V| \ll \Delta/e$, we may approximate the function $F(E)$ [see Appendix~\ref{app:F_E}] to obtain an expression for the elastic tunneling rate
\begin{equation}\label{eqn:elas_off}
 \Gamma^{\rm off}_{\rm el}(eV)\approx\frac{2R_{K}}{R_T} \frac{\gamma_{D}   eV }{ h }  \coth\left(\frac{eV}{2k_{B}T_{N}}\right).
\end{equation}
Considering this rate in the absence of bias voltage we obtain a simple expression for the jump rate in the off-state
\begin{equation}\label{eqn:chioff}
\chi^{\rm off}=\Gamma^{\rm off}_{\rm el}(0)  \approx \frac{4 R_{K}}{R_{T}} \frac{\gamma_{D} k_{\rm B}T_{N}}{h },
\end{equation}
which depends only on the subgap density of states through $\gamma_{D}$, the tunnel resistance $R_{T}$, and the normal-metal electron temperature $T_{N}$.

For a transmon and a QCR with parameters of Table~\ref{tab:parameters}, the charge dispersion between the lowest levels is $\xi/2\pi\approx$~\SI{6}{\kilo\hertz}. In the low-temperature regime of $T_N=$~\SI{10}{\milli\kelvin}, the jump rate $\chi^{\rm off}$  given by Eq.~(\ref{eqn:chioff}) is smaller than the charge dispersion, implying that the off-state fluctuations are in the slow-jumping regime and may increase the transmon dephasing rate by approximately $\chi^{\rm off}=1$~kHz. Hence, in the design of the QCR, one may wish to ensure that in the off-state the jump rate is much higher than the transmon charge dispersion. This goal can be achieved by increasing the value of the ratio $E_{J}/E_{C}$ to near \num{100}, thus diminishing the charge dispersion to approximately near \SI{}{Hz}~\cite{Koch07}. Furthermore, the charge dispersion of a transmon between its higher levels, say between the second and the first excited states, can be of the order of \SIrange{0.1}{250}{\kilo\hertz}. Thus, in the manifold of the highly excited states, the off-state telegraph noise could be in the slow-jumping regime, inducing additional dephasing equal to the jump rate $\chi^{\rm off}$. It is also possible to reduce the off-state jump rate by decreasing $\gamma_D$, see Eq.~\eqref{eqn:chioff}. This would also lower the effective temperature, see Sec.~\ref{sec:operation}, and the off-state relaxation rate, Eq.~\eqref{gamma_off}.

For a conventional flux qubit the charge dispersion can be large, ranging from \SI{50}{\kilo\hertz} to a few \SI{100}{\mega\hertz}~\cite{Stern14, Bal15}. However, for  capacitively shunted flux qubits~\cite{Yan16} with parameters of Table~\ref{tab:parameters}, the charge dispersion given by Eq.~\eqref{eqn:flux1Dexpand} becomes of the order of \SI{1}{\hertz}, bringing the device into the fast-jumping regime in both operation regimes, with negligible additional dephasing from the charge fluctuations. In summary, the QCR-induced temporal charge fluctuations may not introduce significant dephasing in properly designed transmons and capacitively shunted flux qubits. In the case where the device is susceptible to the charge noise of the QCR, its decoherence is likely to be dominated by the ubiquitous charge noise in the supporting materials and their interfaces.

\section{Lamb and ac Stark shifts} \label{sec:Lamb}
In addition to state transitions and temporal frequency fluctuations, a quantum-circuit refrigerator also deterministically shifts the energy levels of the system to which it is coupled. This effect can be regarded as a frequency renormalization by the electromagnetic environment formed by the quantum-circuit refrigerator. Thus, the shifts are referred to as the Lamb and the ac Stark shift~\cite{Lamb_1947, Bethe_1947, Autler_1955, WallsMilburn}. The Lamb shift is the contribution by the zero-point fluctuations of the environment, that is, the environment shifts the energy levels despite being in a vacuum state. The ac Stark shift is the contribution by the excitations of the environment and it thus depends on the temperature of the environment. An anharmonic system experiences both shifts, in contrast to a harmonic system, in which no no ac Stark shift is present~\cite{Carmichael}. Here, we generalize the derivation of Ref.~\cite{Silveri_2019} for a harmonic system to a generic anharmonic system and study the shifts in transmon devices and capacitively shunted flux qubits.

We calculate the shifts $\delta E_m$ in the energies $E_m$ of the quantum network shown in Fig.~\ref{fig:QCR_circuit_gen} using the second-order time-independent perturbation theory. The perturbation is the tunneling Hamiltonian $\hat H_{T}$ of Eq.~\eqref{eq:HT}, and hence the corresponding shift in the energy $E_\eta^\textrm{tot}$ of the total system including the microscopic degrees of freedom can be expressed as
\begin{equation}
  \delta E_{\eta}^\textrm{tot}=\sum_{\eta'\neq \eta}\frac{|\braket{\eta^\prime| \hat H_{T}|\eta}|^2}{E_\eta-E_{\eta^\prime}}, \label{eq.eshift}
\end{equation}
where $\ket{\eta}=\ket{q, m_q, \ell, k}$ denotes an unperturbed eigenstate of the total system, $\ket{q}$ refers to the charge state of the normal-metal island, $\ket{m_q}$ is an eigenstate of the quantum circuit, $\ket{\ell}$ denotes an electron state of the normal-metal island, and $\ket{k}$ is a quasiparticle state in the superconductor with the combined electron-quasiparticle energy $E_{\ell k}=\varepsilon_{\ell}+\epsilon_k$. The total energy of the state $\ket{\eta}$ is $E_\eta^\textrm{tot}=E_{N} q^2+E_m+E_{\ell k}$. By separating the electron-quasiparticle tunneling and quantum-circuit transitions in the perturbation Hamiltonian $\hat H_{T}$ similar to Eq.~\eqref{eq:HT}, we express the energy shift as
\begin{align}
  \delta E_{\eta}^\textrm{tot}=& \sum_{m', \ell', k'}  |M_{mm'}|^2 \notag \\ & \times  \bigg[ \frac{|\braket{\ell' k'| \hat \Theta^\dag|\ell k}|^2}{E_{\ell^{} k^{}}-E_{\ell^\prime k^\prime}-E_{N}(1-2q)+\tilde E_{mm'}}\notag \\ &\quad  +\frac{|\braket{\ell' k'|\hat \Theta|\ell k}|^2}{E_{\ell^{} k^{}}-E_{\ell^\prime k^\prime}-E_{N}(1+2q)+\tilde E_{mm'}} \bigg], \label{eq.eshift.q}
\end{align}
where $M_{m'm}$ denotes the matrix element of Eq.~\eqref{eqn:Mmm} and $\hbar \tilde E_{mm'}=E_m-E_{m'}$ the energy difference between the charge shifted eigenstates $\ket{m_q}$ and $\ket{m'_{q\pm 1}}$ of the quantum circuit.

Note that in Eq.~\eqref{eq.eshift.q}, there is no summation over the final charge states of the normal-metal island since the perturbation connects just the adjacent charge states $q\to q\pm 1$. However, we can average over the initial charge states, similar to Ref.~\onlinecite{Silveri_2017}, since we are just interested in the energy level shift in the quantum circuit regardless of the internal state of the quantum-circuit refrigerator. Typically, the capacitance of the normal-metal island is so large that it makes the charging energy $E_{N}$ the smallest energy scale in the setup. Thus, we can consider only the leading order correction in $E_{N} q $, the contribution of which vanishes for a symmetric charge distribution of initial charge states~\cite{Silveri_2017}.

A quantum-circuit refrigerator consists of two consecutive superconductor--insulator--normal-metal junctions. Assuming that the temperatures and conductances of the junctions are nominally identical, the other junction gives an identical energy shift except that the voltage bias $eV$ is reversed. Thus, the combined energy level shift is given by
\begin{align}
    \delta &E_m=\frac{\hbar}{2\pi} \sum_{\tau=\pm 1}\sum_{m'} \frac{2 R_{K} }{ R_{T}} |M_{mm'}|^2 \  \\
     & \times \bigg\{ \frac{1}{h} \textrm{PV}\iint_{-\infty}^{\infty} d{\epsilon_k} d{\varepsilon_\ell} \frac{n_{S}(\epsilon_k)\left[1-f(\epsilon_k)\right]f(\varepsilon_{\ell})}{\varepsilon_{\ell}-\epsilon_k-E_{N}+E_{mm'}+\tau eV} \notag \\
                         & \quad+\frac{1}{h} \textrm{PV}\iint_{-\infty}^{\infty} d{\epsilon_k} d{\varepsilon_\ell} \frac{n_{S}(\epsilon_k)f(\epsilon_{k})\left[1-f(\varepsilon_\ell)\right]}{\epsilon_k-\varepsilon_{\ell}-E_{N}+E_{mm'}-\tau e V} \bigg \}, \notag
\end{align}
where $\textrm{PV}$ means the Cauchy principal value integration. By utilizing the normalized tunneling functions $F(E)$ of Eq.~\eqref{eqn:F} and their properties, the energy level shift reduces to a compact form
\begin{align}
    \delta E_m  = & -\hbar \sum_{\tau=\pm 1}\sum_{m'}  \frac{2 R_{K}}{R_{T}} |M_{mm'}|^2 \notag \\ &\qquad\quad  \times \textrm{PV} \int_{-\infty}^\infty \frac{d \omega}{2\pi} \frac{F(\tau eV + \hbar\omega-E_{N})}{\omega+\omega_{mm'}} \notag \\
    = & -\hbar \sum_{m'}  \textrm{PV} \int_{-\infty}^\infty \frac{d \omega}{2\pi} \frac{\Gamma_{mm'}(\omega)}{\omega+\omega_{mm'}}\, , \label{eq:shifts}
\end{align}
where $\omega_{mm'}=\tilde E_{mm'}/\hbar$ and $\Gamma_{mm'}(\omega)$ is the transition rate of Eq.~\eqref{eqn:Gamma} from the state $\ket{m}$ to the state $\ket{m'}$ assuming that their energy separation is $\hbar \omega$,  not $\hbar \omega_{mm'}$.

\begin{figure}[t]
  \centering
  \includegraphics[width=1\linewidth]{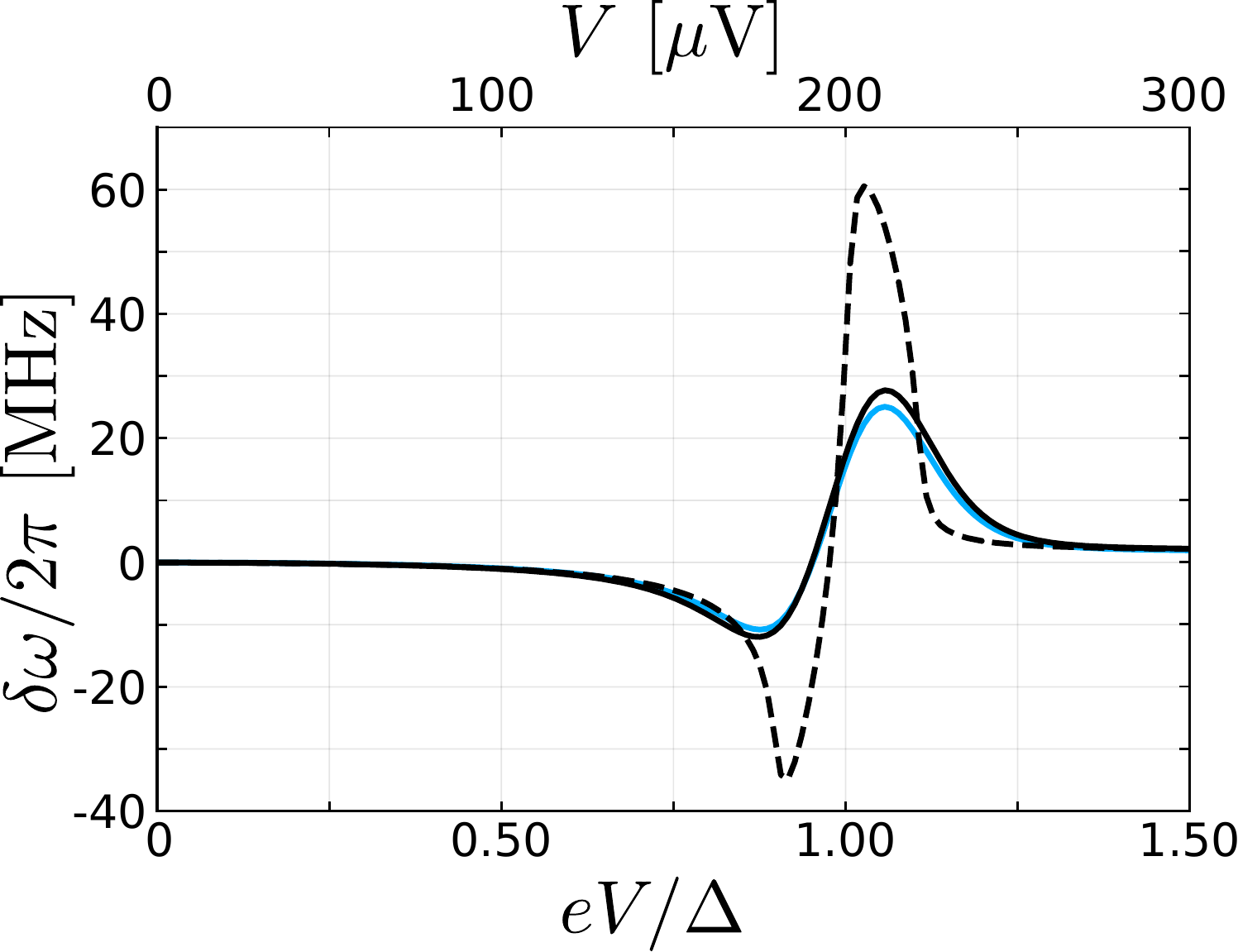}
  \caption{\label{fig:lamb} Energy level shifts $\delta \omega$ as a function of the bias voltage $V$ for a harmonic oscillator (black) and for the 1-2 transition in a transmon qubit (blue) as given by Eqs.~\eqref{eq:oshift} and \eqref{eq:anshift} at the electron temperature $T_{N}=$~\SI{100}{\milli\kelvin} and the anharmonicity $U/2\pi=$~\SI{0.303}{\giga\hertz} in accordance to Fig.~\ref{fig:06}. For comparison, the dashed line correspond to the oscillator shift from Eq.~\eqref{eq:oshift} at the lower QCR electron temperature $T_{N}=$~\SI{10}{\milli\kelvin}.  The QCR and transmon parameters are provided in Table~\ref{tab:parameters}. }
\end{figure}

If the QCR is coupled to a harmonic oscillator with the energy splitting $\omega_{mm'}=\omega_0(m-m')$ and with the matrix elements $|M_{mm}|^2=1-(1+2m) \alpha^2 \rho$, $|M_{m,m-1}|^2=m \alpha^2 \rho$, and $|M_{m,m+1}|^2=(m+1) \alpha^2 \rho$ in the leading order of the interaction parameter $\rho$ of Eq.~\eqref{eqn:rho}, then the energy level shift $\delta E_{m+1}-\delta E_m=\hbar \delta \omega^{\rm osc}$ reduces to
\begin{align}
    \delta \omega^{\rm osc}=-\textrm{PV} \int_0^\infty\frac{d\omega}{2\pi}  \left[\frac{\gamma(\omega)}{\omega+\omega_0}+\frac{\gamma(\omega)}{\omega-\omega_0}-\frac{2 \gamma(\omega)}{\omega}\right].\label{eq:oshift}
\end{align}
This shift referred to as the dynamic Lamb shift~\cite{Silveri_2019} since it vanishes with $\omega_0\to 0$. It is dependent only on the coupling strength to the environment through $\gamma(\omega)$, which is defined as in Eq.~\eqref{eqn:gamma_T}. Each energy eigenstate experiences energy shifts by the elastic tunneling and excitation- and relaxation-inducing photon-assisted tunneling 
which combined, result in the energy shift of Eq.~\eqref{eq:oshift}.

A transmon device and a capacitively shunted flux qubit are both weakly anharmonic oscillators. Thus, their energy shift characteristics are close to that of a harmonic oscillator of Eq.~\eqref{eq:oshift}. For concreteness, we consider here an anharmonic system with energy level structure $\omega_m=m\omega_0-\frac{U}{2}m(m-1)$, with the anharmonicity $U$ producing energy level splittings $\omega_{m+1,m}=\omega_0-U m$. Furthermore, we approximate the matrix elements by those of a harmonic oscillator, see the text before Eq.~\eqref{eq:oshift}. For a transmon device and for a capacitively shunted flux qubit, this is a reasonable and accurate approximation, as justified in Figs.~\ref{fig:trme} and~\ref{fig:CSFQ_me_EJdEC}. For simplicity, we fix the electron temperature $T_{N}=$~\SI{100}{\milli\kelvin} such that $\hbar U\ll k_{B}T_{N}$. In addition to the matrix elements, the transition rates $\Gamma_{m,m'}$ depend on the transition energy. In a harmonic oscillator, the transition energy is independent of the number of excitations, but in an anharmonic system this is not the case in general. However, since the transition rates $\Gamma_{m,m'}(\omega)$ vary on the energy scales $k_{B}T_{N}$, we may ignore in Eq.~\eqref{eq:shifts} this difference in the excitation and relaxation rates by the anharmonic $\hbar U \ll k_{B}T_{N}$ energy,
which finally yields for $\delta E_{m+1}-\delta E_m=\hbar \delta \omega^{\rm an}_{m+1,m}$
\begin{align}
    \delta \omega^{\rm an}_{m+1,m}=-\textrm{PV}\!\int_0^\infty\!\frac{d\omega}{2\pi} \left[\frac{\gamma(\omega)}{\omega+\omega'_m} +\frac{\gamma(\omega)}{\omega-\omega'_m}-\frac{2\gamma(\omega) }{\omega}\right]\!,\label{eq:anshift}
\end{align}
where $\omega'_m=\omega_0-Um$. With the realistic QCR parameters, the shifts lie in the~10-MHz range as visualized in Fig.~\ref{fig:lamb} and measured in Ref.~\cite{Silveri_2019}.

In addition to the average value considered here, the ac Stark shift results in also fluctuations in the energies of the quantum circuit, and hence dephasing. These energy fluctuations arise from the fluctuations of the photon number about its mean in the thermal state of the environment. However, since the energy shifts are proportional to the QCR-induced decay rate, we may neglect them in the off-state. In the on-state on the other hand, dephasing is not of great importance since the quantum circuit is quickly relaxing very close to its ground state, and consequently loses any phase information in any case.

\section{QCR in circuit QED}\label{sec:QCR_tr_res}
\begin{figure}[tb]
  \centering
  \includegraphics[scale=0.42]{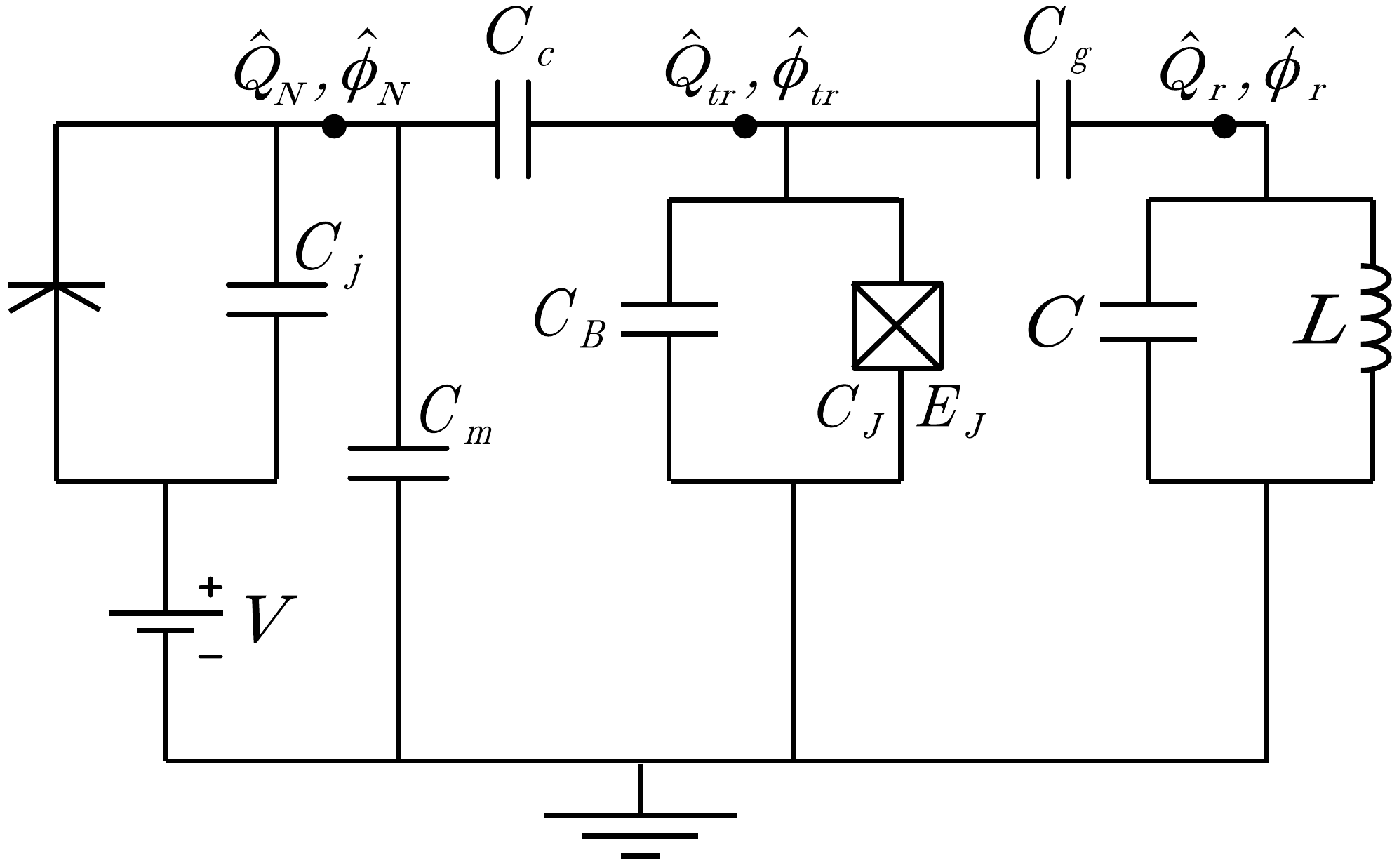}
  \caption{\label{fig:07} Circuit diagram a transmon coupled to a QCR through the capacitance $C_c$  and to a resonator through the capacitance $C_g$, cf. Fig.~\ref{fig:QCR_circuit_gen}.}  
\end{figure}

In the previous sections, we have studied in detail the case of a qubit coupled to a QCR. In circuit QED~\cite{Blais04, Wallraff04}, however, superconducting qubits are coupled to a resonator in the so-called dispersive regime to enable readout of the qubit state. Alternatively, very weakly anharmonic transmons can be used to manipulate the quantum states of cavities~\cite{Heeres15, Krastanov15}.
In this section, we therefore consider two possible configurations consisting of a QCR, a transmon, and a resonator. In the first case, the transmon is capacitively coupled to the QCR by capacitance $C_c$ and to a resonator by capacitance $C_g$, see Fig.~\ref{fig:07}. In the second configuration, we exchange the position of the transmon and the resonator in the quantum circuit, so that the resonator is directly coupled to the QCR.

In the notation of Eqs.~(\ref{eqn:H0}) and~(\ref{eqn:HCp}), with the identifications $\hat Q_\textrm{tr}=\hat Q_0$, $\hat Q_{r} = \hat Q_1$, and $\hat \varphi=2\pi\hat \phi_0/\Phi_0$  the Hamiltonian of the QCR-transmon-resonator system is given by
\begin{align}
&\hat H'_0 = \hat H'_{C}+\hat H_\phi, \notag\\
& \hat H'_{C}=\frac{\hat Q_{tr}^2}{2 \widetilde{C}_{0}} + \frac{\hat Q_{r}^2}{2 \widetilde{C}_M} +\kappa_1 \hat Q_{tr}  \hat Q_{r} + \frac{\hat Q_{N}^2}{2 C_{N}}, \notag \\
&\hat H_\phi=\frac{\hat \phi_{r}^2}{2L} - E_J\cos \hat \varphi,  \label{eqn:H_QCR-tr-res}
\end{align}
where the relations of capacitances $\widetilde{C}_0$ and~$\widetilde{C}_M$, 
and of the coupling parameter $\kappa_1$ (an inverse capacitance) to the circuit capacitances are explicitly given in Appendix~\ref{app:inverseC} and the charge offsets are removed.
In the QCR-resonator-transmon configuration, we simply exchange the transmon and the resonator, and hence the Hamiltonian has a similar form.

To proceed, we apply the standard procedure discussed in Ref.~\cite{Koch07}. In its second-quantization representation, after the rotating-wave approximation is applied, the above Hamiltonian is given by
\begin{align}
\hat H_0&\approx\frac{\hat Q_{N}^2}{2C_{N}}+\hbar \omega_r \hat a^\dag \hat a+ \sum_{j=0} \hbar \omega_j \ket{j}\bra{j} \notag\\
&+\sum_{j=0} \hbar \lambda_j \left( \hat a \ket{j+1}\bra{j} + \hat a^\dag \ket{j}\bra{j+1}\right),
\label{eqn:H2nd_QCR-res-tr}
\end{align}
where $\omega_r$ is the resonator frequency, $\hat a^\dagger$ ($\hat a$) is the creation (annihilation) operator for photons in the resonator,
$\left| j \right\rangle$ is the $j$th eigenstate of the transmon with energy $\hbar\omega_j$, and $\lambda_j=\lambda \sqrt{j+1}$ are the leading-order coupling constants. In terms of the parameters in Eq.~(\ref{eqn:H_QCR-tr-res}), these quantities are given by~\cite{Koch07}
\begin{align}
    \omega_r & = 1/\sqrt{L\widetilde{C}_M}, \\
    \hbar \omega_j & \approx \sqrt{8E_C E_J}\left(j+\frac{1}2\right) - \frac{E_C}{12} \left(6j^2+6j+3\right), \\
    \lambda & =-\kappa_1 \left[-2e i \left(\frac{E_J}{8E_C} \right)^{\frac{1}{4}} \frac{1}{\sqrt{2}}\right] \left( -i\sqrt{\frac{\hbar \omega_r \widetilde{C}_M}{2}}\right) \notag\\
    &=\kappa_1 e \left(\frac{E_J}{8E_C} \right)^{\frac{1}{4}} \sqrt{\hbar \omega_r \widetilde{C}_M},
\end{align}
where $E_C=e^2/(2\widetilde{C}_0)$. To approximately diagonalize the Hamiltonian, we utilize a unitary Schrieffer-Wolff transformation \cite{Koch07} $e^{\hat S}$ with $\hat S=\sum_j \frac{\lambda_j}{\Delta_j} (\hat a  \ket{j+1}\bra{j}- \hat a^\dag  \ket{j}\bra{j+1})$, where $\Delta_j=\omega_{j+1}-\omega_j-\omega_r$.
In the dispersive limit $|\lambda_j/\Delta_j| \ll 1$, keeping up to next-to-leading terms and truncating the transmon part to the two lowest levels, we have
\begin{equation}\label{eq:transH}
e^{\hat S} \hat H_0 e^{-\hat S}\approx\frac{\hat Q_{N}^2}{2C_{N}}+\frac{\hbar\widetilde \omega_{01}}{2}\hat \sigma_z+\hbar (\widetilde \omega_r+\chi\hat\sigma_z) \hat a^\dag  \hat a,
\end{equation}
where the qubit and the resonator frequencies, given by $\widetilde \omega_{01}=\omega_1-\omega_0+\chi_{01}$ and $\widetilde \omega_r=\omega_r-\chi_{12}/2$, respectively, are shifted by the dispersive shifts $\chi_{01}=\lambda^2/\Delta_0$, $\chi_{12}=\lambda_1^2/\Delta_1$, and $\chi=\chi_{01}-\chi_{12}/2$.

Next, we calculate the transition matrix elements. To this end, we label the eigenstates of the transformed Hamiltonian in Eq.~\eqref{eq:transH} as $\left| \widetilde m \right\rangle = \ket{\tilde j, \widetilde m_r}$, with ${m}_r = 0,\,1,\,2,\, \ldots$ and $j=0,\, 1$. Note that the NIS tunneling Hamiltonian in Eq.~(\ref{eqn:HTp}), and hence the matrix elements in Eq.~(\ref{eqn:Mmm}), are affected by the unitary transformation $e^{\hat S}$.
We use the Baker-Hausdorff identity $e^{\hat X} \hat Y e^{-\hat X} = \hat Y + [\hat X,\hat Y]/1!+[\hat X,[\hat X,\hat Y]]/2! + \ldots$ and keep
terms up to the first order in $\frac{\lambda_j}{\Delta_j}$,  since all higher-order terms result in only a small correction to the transition matrix elements. As a result, we have
\begin{equation}
e^{ \hat S} e^{-i\frac{e}{\hbar} \alpha \hat \phi_{0}} e^{- \hat S} \approx e^{-i\frac{e}{\hbar} \alpha \hat \phi_{0}}+[ \hat S,e^{-i\frac{e}{\hbar} \alpha \hat \phi_{0}}],
\end{equation}
where $\hat \phi_0=\hat \phi_\textrm{tr}$.
As in Sec.~\ref{sec:Transmon}, we treat the transmon as a weakly anharmonic oscillator.
The resulting transition matrix elements for the QCR-transmon-resonator (QTR) configuration are given by
\begin{align}
&M^{\rm QTR}_{j m_r, j' m'_r} = \bra{\tilde j', \widetilde m'_r} e^{ \hat S}e^{-i\frac{e}{\hbar} \alpha \hat \phi_{\rm tr}} e^{-\hat S} \ket{\tilde j, \widetilde m_r} \approx M^\text{ho}_{ j  j'} \delta_{ m_r  m'_r } \notag \\
&+\sqrt{ m_r} \delta_{ m'_r,  m_r-1} \left (\frac{\lambda_{ j'-1}}{\Delta_{ j'-1}} M^\text{ho}_{ j,  j'-1}- \frac{\lambda_{ j}}{\Delta_{ j}} M^\text{ho}_{ j+1,  j'}\right) \notag \\
&-\sqrt{ m_r+1} \delta_{ m'_r,  m_r+1} \left(\frac{\lambda_{ j'}}{\Delta_{ j'}} M^\text{ho}_{ j,  j'+1}- \frac{\lambda_{ j-1}}{\Delta_{ j-1}} M^\text{ho}_{ j-1,  j'}\right),
\end{align}
where the harmonic-oscillator matrix elements $M^\text{ho}_{ j  j'}$ are given in Eq.~(\ref{eq:Mho}).
For example, the matrix element for the relaxation of the resonator assumes the form
\begin{equation} \label{eqn:QCR-tr-res-me}
\left| M^{\rm QTR}_{01,00} \right|^2=\left(\frac{\lambda}{\Delta_0}\right)^2 |M^\text{ho}_{10}|^2.
\end{equation}
The prefactor $\left(\lambda/\Delta_0\right)^2$ in this equation is fully analogous to the well-known Purcell suppression of the decay rate for a qubit dispersively coupled to a resonator~\cite{Blais04}, although here the roles of the qubit and the resonator are exchanged.

For the QCR-resonator-transmon configuration (QRT), similarly we have
\begin{equation}
e^{ \hat S} e^{-i\frac{e}{\hbar} \alpha \hat \phi_r} e^{- \hat S} \approx e^{-i\frac{e}{\hbar} \alpha \hat \phi_r}+[ \hat S,e^{-i\frac{e}{\hbar} \alpha \hat \phi_r}].
\end{equation}
Here, we also use Eq.~(\ref{eqn:Mmm}) to obtain the transition matrix elements
\begin{align}
&M^{\rm QRT}_{ j  m_r  j'  m'_r}= \bra{\tilde j', \widetilde m'_r} e^{ \hat S}e^{-i\frac{e}{\hbar} \alpha \hat \phi_r} e^{-\hat S}\ket{\tilde j, \widetilde m_r} \approx M^\text{ho}_{ m_r  m'_r}\delta_{ j j' } \notag \\
&+\frac{\lambda_{ j}}{\Delta_{ j}} \delta_{ j', j+1} \left(\sqrt{ m'_r+1} M^\text{ho}_{ m_r,  m'_r+1}-\sqrt{ m_r} M^\text{ho}_{ m_r-1,  m'_r} \right) \notag \\
&-\frac{\lambda_{ j-1}}{\Delta_{ j-1}} \delta_{ j', j-1}
\left(\sqrt{ m'_r} M^\text{ho}_{ m_r,  m'_r-1}-\sqrt{ m_r+1} M^\text{ho}_{ m_r+1,  m'_r}\right)\!.
\end{align}
With the matrix elements, we can calculate the transition rates $\Gamma_{ j  m_r  j'  m'_r}(V)$ by Eq.~(\ref{eqn:Gamma}).

\begin{figure}[tb]
  \centering
  \includegraphics[width=1.0\linewidth]{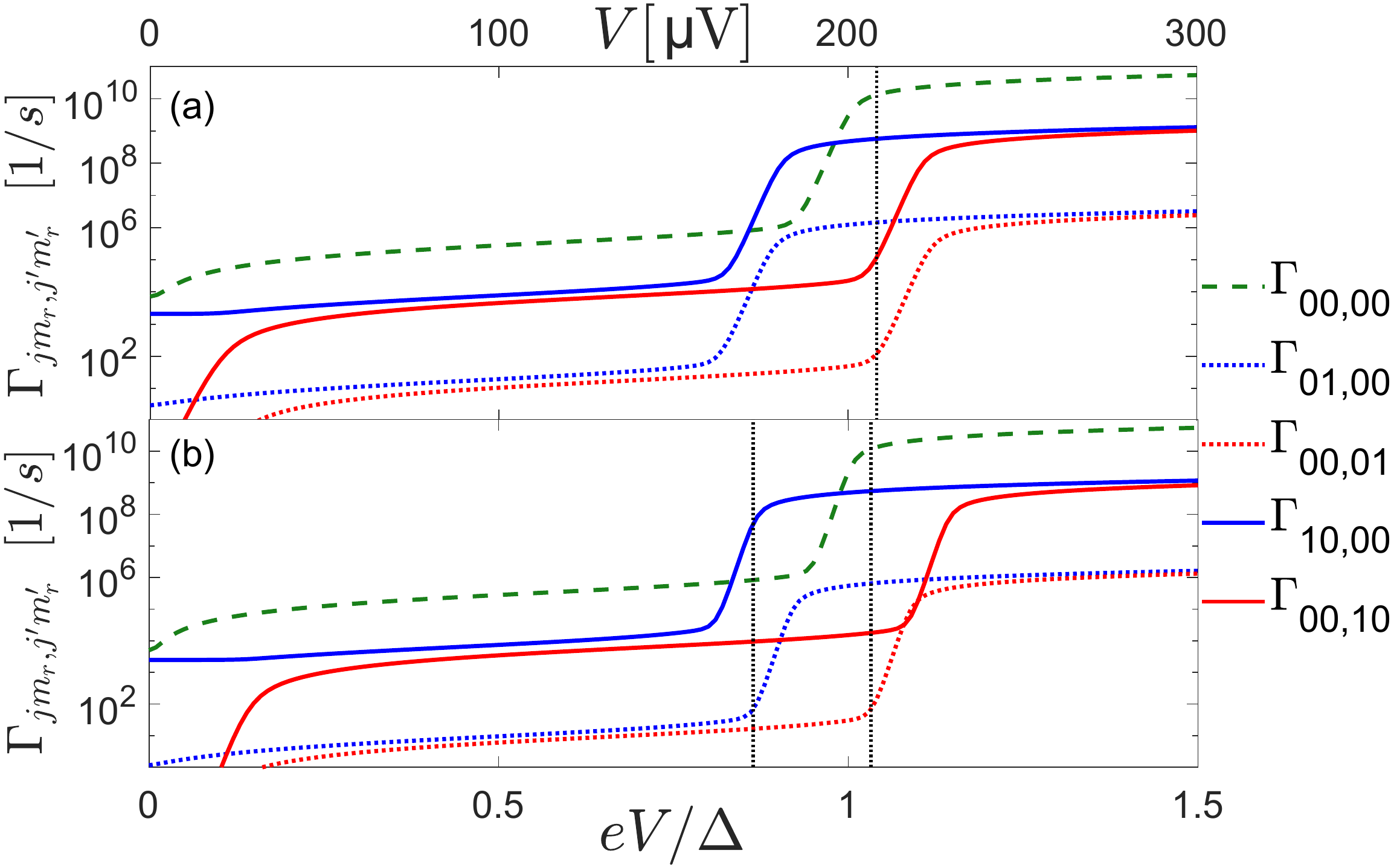}
  \caption{\label{fig:08} Transition rates in the QCR-transmon-resonator system $\Gamma_{ j  m_r,  j'  m'_r}(V)$ as functions of the bias voltage $V$. Parameters are taken from (a) Ref.~\onlinecite{Schoelkopf_2016_Sciences} where the resonator has a transition frequency of~\SI{5.45}{\giga\hertz} and the transmon operates at~\SI{4.88}{\giga\hertz}, and from (b) Ref.~\onlinecite{Schoelkopf_2017_Natphys} with the resonator and transmon frequencies~\SI{4.07}{\giga\hertz} and~\SI{6.67}{\giga\hertz}, respectively. The vertical black dotted lines denote the operation voltages. In (a) $eV=eV_{\rm max} \approx 1.04 \, \Delta$ and in (b) $eV=eV_{\rm max} \approx 1.03 \, \Delta$ (right) and $eV=eV_{\rm max}-2\hbar \omega_{10} \approx 0.86 \, \Delta$~(left). We use Eq.~(\ref{eq:eVmax}) to calculate $eV_{\rm max}$ with $\omega_{10}/2\pi$ as the frequency of the resonator. The QCR parameters are as in Table~\ref{tab:parameters}. The electron temperatures $T_N$ are (a)~\SI{20}{\milli\kelvin} and (b)~\SI{15}{\milli\kelvin}.}
\end{figure}

Here, we discuss a possible application of the QCR in experiments with high-quality resonators. For example, in the setup of Ref.~\onlinecite{Schoelkopf_2016_Sciences}, two high-quality three-dimensional (3D) coaxial cavities are used as quantum memories. The cavities are coupled to a transmon ancilla on a sapphire chip. The transmon is used for manipulating the cavity states. In this type of experiments, it is not advantageous to wait for the spontaneous relaxation due to the long lifetimes of the two cavities of roughly \SI{1.5}{\milli\second} and~\SI{3}{\milli\second}. In fact, in the experiment, a so-called four-wave mixing processes is implemented to initialize both cavities in roughly \SI{400}{\micro\second}.

In contrast, we consider the QCR-transmon-resonator configuration which is more convenient to realize than coupling the QCR to the 3D cavities. Figure~\ref{fig:08}(a) shows our results on the QCR-induced transition rates using parameters from Ref.~\onlinecite{Schoelkopf_2016_Sciences}. At the on-voltage $eV=eV_{\rm max} \approx 1.04 \, \Delta$, obtained from Eq.~(\ref{eq:eVmax}) with $\omega_{10}$ as the frequency of the cavity, the relaxation rate of the transmon $\Gamma_{10,00}$ is roughly~\SI{5.7e8}~$1/\textrm{s}$. The resonator transition matrix elements are multiplied by a factor of $(\lambda/\Delta_0)^2 \approx$~\num{0.0024}, resulting in a smaller relaxation rate of $\Gamma_{01,00} \approx$~\SI{1.4e6}~$1/\textrm{s}$. The corresponding reset time for the resonator is approximately $T_{10\%}=\ln(10)/\Gamma_{01,00} \approx$~\SI{1.6}{\micro\second},
much shorter than that of the four-wave mixing protocol. Note that the QCR simultaneously resets both the cavity and the transmon, which may be useful in this type of experiments to ensure that initially both of these systems are in their ground states.

As in Sec.~\ref{sec:qubits}, we assume that the reset fidelity is limited by the excitations due to the QCR and that the temperature $T_N$ is below the cross-over temperature. Thus the infidelity $1-{\cal F}_r$ for the resonator is
$\Gamma_{00,01}/\Gamma_{01,00} \approx$~\num{7.6e-5}, and for the transmon, we have $\Gamma_{00,10}/\Gamma_{10,00}\approx$~\num{1.9e-4}.

The speed of the four-wave mixing protocol varies from setup to setup. For example, Ref.~\onlinecite{Schoelkopf_2017_Natphys} reports a reset time $T_{10 \%}$ of about \SI{2.5}{\micro\second}. For QCR, at $eV=eV_{\rm max} \approx 1.03 \, \Delta$ calculated by Eq.~(\ref{eq:eVmax}) with $\omega_{10}$ as the cavity frequency, we obtain in Fig.~\ref{fig:08}(b) a reset time of $T_{10 \%}\approx~\SI{3.6}{\micro\second}$. Interestingly, with the parameters of Ref.~\cite{Schoelkopf_2017_Natphys} there is another possibly useful operation voltage indicated in Fig.~\ref{fig:08}(b): at $eV=eV_{\rm max}-2\hbar \omega_{10} \approx 0.86 \, \Delta$ the transmon reset time is $T_{10 \%}=\ln(10)/\Gamma_{01,00} \approx$~\SI{0.07}{\micro\second}, while the QCR barely affects the cavity, since both the up and down rates are less than~\SI{100}{\hertz}. Therefore, one can reset the transmon without disturbing the cavity. This operation point is possible because the  cavity frequency is well below the transmon frequency, and hence its decay is not activated at such a low operation voltage.

In summary, the QCR provides an alternative reset protocol which, depending on the details of the setup, may be faster than the existing protocols in experiments with high-quality resonators. The QCR has an additional benefit, namely, it may be used to reset also the ancilla transmon.

\section{Conclusions}
\label{sec:conclusions}
We have provided a general theoretical framework to study a quantum-circuit refrigerator coupled to a network composed of possibly non-linear quantum electric circuit elements. As example cases of the network, we studied a transmon qubit, a capcitively shunted flux qubit, and a qubit-resonator system. By applying our general theoretical results for several different experimentally realized quantum circuits, we concluded that the QCR seems a promising candidate to reset state-of-the-art qubits quickly and accurately to their ground states. Namely, with typical experimental parameters given in Table~\ref{tab:parameters}, we obtain that the excited-state population in a transmon or in a flux qubit decreases by an order of magnitude in~\SI{1.5}{\nano\second} or in~\SI{5.5}{\nano\second}, respectively, with a steady-state infidelity of $5\times 10^{-5}$ (see Table~\ref{tab:key_results}).

Interestingly, the qubits tend to couple naturally more strongly to the QCR than to 50-$\Omega$ transmission lines, which provides additional flexibility in the design parameters. Namely, the minimum QCR-induced decay rate on the qubit is roughly $\bar\gamma\gamma_D\simeq \gamma_D/(1\textrm{ ns})$ in the case of maximal capacitive coupling. This may be decreased, for example, by reducing in fabrication the coupling capacitance between the QCR and the qubit, by increasing the tunneling resistance of the QCR tunnel junction, or by coupling the QCR to the qubit through a resonator as introduced in this work.
For example, for a typical bare qubit decay time of $100$~$\mu$s and a Dynes parameter of $\gamma_D=10^{-5}$, we obtain QCR-coupled-qubit decay time of $T_1=75$~$\mu$s provided that we slightly reduce the QCR-qubit coupling capacitance such that the QCR-induced decay rate in the off state is $1/(300\textrm{ }\mu\textrm{s})$. Thus if properly designed, the QCR does not significantly increase the gate or readout error, but will allow for a quick reset of the quantum circuit since the decay rate of the qubit goes up by a factor of roughly $1/\gamma_D$ when the QCR is turned on.

In the future, we aim to extend our theoretical model into a master equation taking explicitly into account the charge fluctuations, population decay, and frequency shifts induced by the QCR. Such technique enables us to dynamically model a quantum network subject to quantum-circuit refrigeration, including the effect of the QCR on qubits in the vicinity of that being initialized. Furthermore, we are working on an experimental realization of a QCR-coupled qubit to verify the theoretical predictions laid out in this paper.

\begin{acknowledgments}
We acknowledge Hermann Grabert, Jani Tuorila, Vasilii Sevriuk, and Eric Hyypp{\"a} for useful discussions. This research was financially supported by: European Research Council under Grant No.~681311 (QUESS) and Marie Sk\l{}odowska-Curie Grant No.~795159; Academy of Finland under its Centres of Excellence Program grants No.~312300, and grants Nos.~265675, 305237, 305306, 308161, 312300, 314302, 316551,
316619, 
320086; 
Finnish Cultural Foundation; the Jane and Aatos Erkko Foundation; Vilho, Yrj\"{o} and Kalle V\"{a}is\"{a}l\"{a} Foundation;
the Emil Aaltonen Foundation; 
the Alfred Kordelin Foundation; 
the Technology Industries of Finland Centennial Foundation; and the Internationalization Fund - Cutting-Edge Ideas initiative of Forschungszentrum J\"ulich.
\end{acknowledgments}

\appendix

\section{Approximate formulas for normalized rate of single-electron tunneling} 
\label{app:F_E}

We study the normalized tunneling rate function $F(E)$ of Eq.~(\ref{eqn:F}) assuming $\gamma_D \Delta, \, k_B T_N \ll \Delta$. We consider $E>0$, since for negative arguments we can use the relation $F(-E) = e^{-E/(k_B T_N)}F(E)$~\cite{Silveri_2017}. For energies above the gap, $E>\Delta$, we have [cf. Eq.~(\ref{eq:F_0T})]
\begin{equation}\label{eqn:Fabovegap}
    F(E) \approx \frac{1}{h}\sqrt{E^2-\Delta^2}\, ,
\end{equation}
valid when $E-\Delta \gg \max\{k_B T_N,\,\gamma_D\Delta\}$. Finite temperature corrections to Eq.~\eqref{eqn:Fabovegap} can be found using a Sommerfeld expansion~\cite{Silveri_2017, Hyyppa19}.
For energies below the gap, $E<\Delta$, we can distinguish two cases.
First, for energies sufficiently far from $\Delta$ (in a sense to be clarified below), we can again use a Sommerfeld expansion approach to find an approximation for $F(E)$
to be
\begin{equation}\label{eqn:FlowTN}
F(E) \approx  \frac{1}{h}\frac{\gamma_D  E\Delta}{\sqrt{\Delta^2-E^2}} \frac{1}{1-e^{-E/(k_B T_N)}}\, .
\end{equation}
Compared to the zero-temperature limit in Eq.~(\ref{eq:F_0T}), including the last factor leads to the correct leading-order expression for $T_N>0$ and $E\ll k_B T_N$. 

In practice, we are interested in the regime $k_B T_N \gg \gamma_D \Delta$. Thus,  for energies close to $\Delta$ we take as a starting point the limit $\gamma_D\to0$ and find the thermal activation formula~\cite{Silveri_2017}
\begin{equation}
F(E) \approx  \frac{1}{h}\sqrt{\frac{\pi k_B T_N \Delta}{2}} \textrm{exp}\left( \frac{E-\Delta}{k_B T_N} \right),
\label{eqn:Fthermal}
\end{equation}
which also requires $\Delta-E \gg k_B T_N$.
As the energy $E$ is lowered, the right hand side of Eq.~(\ref{eqn:Fthermal}) becomes exponentially small, while the decrease of the right hand side of Eq.~(\ref{eqn:FlowTN}) is approximately linear in energy. This indicates that at finite $\gamma_D$ and for any given temperature $T_N$, there is a cross-over energy $E_\mathrm{co}$ below which the broadening in the  density of states and the resulting subgap tail cannot be neglected as in Eq.~(\ref{eqn:Fthermal}). This cross-over energy can be found
by equating Eqs.~(\ref{eqn:FlowTN}) and (\ref{eqn:Fthermal}) and solving the resulting equation by iterations. For $k_B T_N \gg \gamma_D \Delta$ we obtain
\begin{align}
E_\mathrm{co} \approx \Delta- k_B T_N \bigg\{ &\ln \left( \frac{\sqrt{\pi}k_B T_N}{\gamma_D \Delta}\right)\notag \\& \quad + \frac{1}{2} \ln \left[ \ln \left( \frac{\sqrt{\pi}k_B T_N}{\gamma_D \Delta} \right)\right]\bigg \}. \label{eqn:Eco}
\end{align}
In summary, Eq.~(\ref{eqn:FlowTN}) is valid for $E<E_\mathrm{co}$ and Eq.~(\ref{eqn:Fthermal}) for $E>E_\mathrm{co}$ and $\Delta-E \gg k_B T_N$.

\begin{figure}[tb]
  \centering
  \includegraphics[width=1\linewidth]{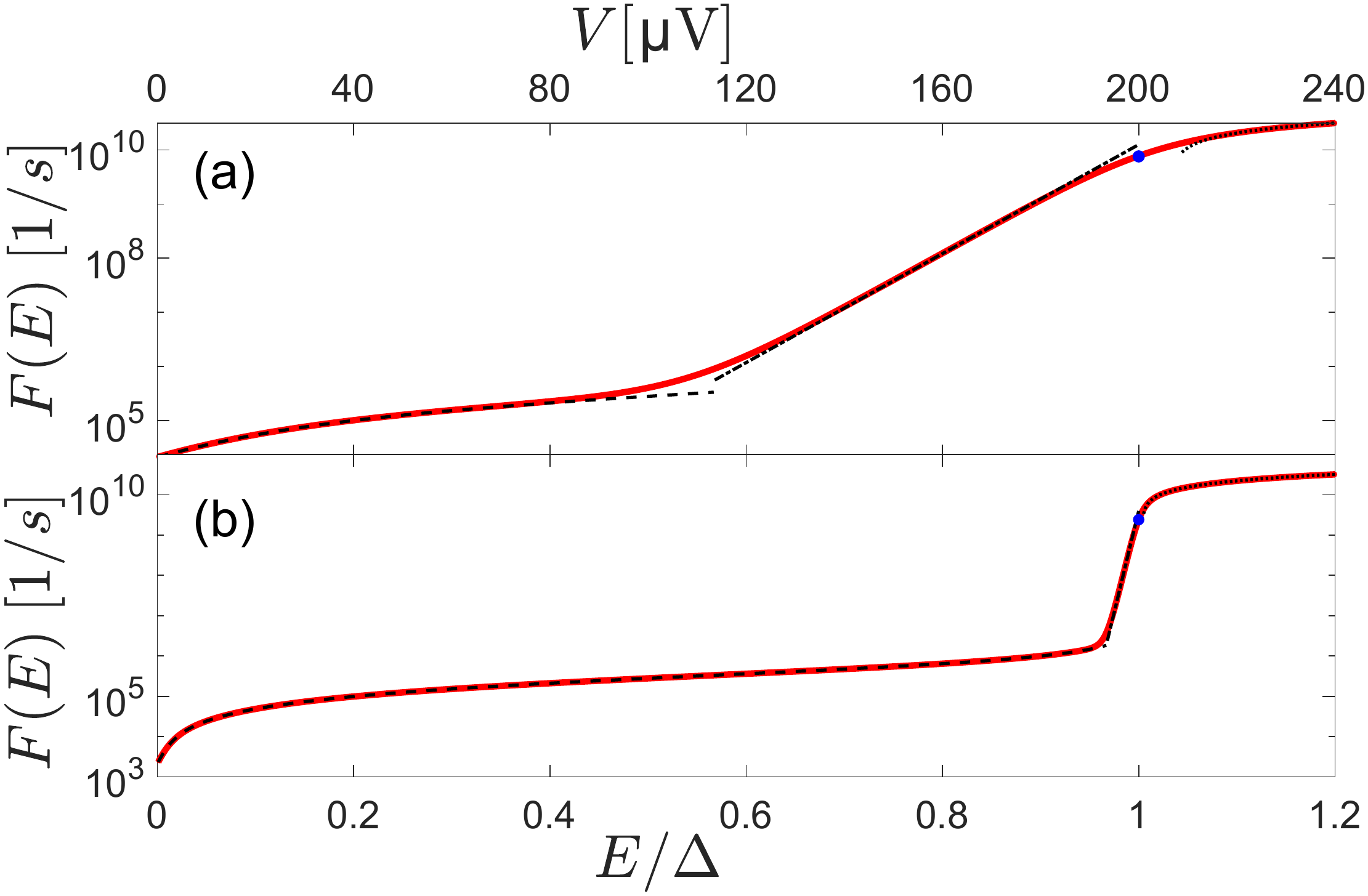}
  \caption{\label{fig:09} Normalized tunneling rate $F(E)$ as a function of the bias energy $E$ at the electron temperature (a)~$k_B T_N/\Delta = 0.043$ and (b)~$k_B T_N/\Delta = 0.0043$. Black dashed lines represent the approximate results from Eq.~(\ref{eqn:FlowTN}) for $E < E_{\rm co}$. Black dash-dotted lines are from Eq.~(\ref{eqn:Fthermal}) for $E_{\rm co} < E < \Delta$. Black dotted lines are from Eq.~(\ref{eqn:Fabovegap}) for $E > \Delta+k_B T_N$. Blue dots are Eq.~(\ref{eqn:Fdelta}) for $E=\Delta$. The Dynes parameter is $\gamma_D = 10^{-5}$. }
\end{figure}

At $E=\Delta$, Eq.~(\ref{eqn:Fthermal}) gives the correct dependence on the temperature but overestimates the numerical prefactor. For a more accurate expression, we again take $\gamma_D \to 0$ in Eq.~(\ref{eqn:F}), change the integration variable from $\varepsilon$ to $\Delta + k_B T_N x$, and neglect terms that are small in $k_B
T_N/\Delta$ to find
\begin{equation}
\label{eqn:Fdelta}
    F(\Delta) \approx  \frac{1}{h}\sqrt{\frac{k_B T_N\Delta}{2}} \int^\infty_0 \frac{dx}{\sqrt{x}}\frac{1}{e^x+1} \, .
\end{equation}
The last integral is a complete Fermi-Dirac integral and can be expressed in terms of a polylogarithm or equivalently via the Dirichlet eta function. Here, we use the Riemann zeta function to write
\begin{equation}
    \int^{\infty}_0 \frac{dx}{\sqrt{x}}\frac{1}{e^x+1} = \sqrt{\pi}\left(1-\sqrt{2}\right)\zeta\left(\frac12 \right) \approx 1.072.
\end{equation}
In Fig.~\ref{fig:09}, we compare the approximate results presented above to numerical calculations of $F(E)$ and find good agreement in all the relevant energy ranges.

\section{Inverse of the capacitance matrix}
\label{app:inverseC}

In this appendix, we find the inverse of the capacitance matrix $\boldsymbol{C}$ of the circuit represented schematically in Fig.~\ref{fig:QCR_circuit_gen} and derive Eq.~\eqref{eqn:HC}. The kinetic energy part of the Lagrangian can be written in the form
\begin{equation} \label{eqn:Cmatrix} \begin{array}{l}
T=\frac{1}{2} \dot{\vec{\Phi}}^T \boldsymbol{C} \; \dot{\vec{\Phi}},\\
\boldsymbol{C} = \left( {\begin{array}{*{20}{c}}
   C_N             & -C_c            & 0  \\
   -C_c            & C_0             & \vec{C}_V^T  \\
   0               & \vec{C}_V             & \boldsymbol{C}_M \\
\end{array}} \right),\;
\vec{\Phi}= \left( {\begin{array}{*{20}{c}}
   \Phi_N  \\
   \Phi_0  \\
   \vec{\Phi}  \\
\end{array}} \right),
\end{array}\end{equation}
where $\vec{\Phi}=(\Phi_1,\Phi_2,\ldots,\Phi_M)^T$, $C_N$ is defined after Eq.~\eqref{eqn:HC}, and $C_0$ is given by the sum of all the capacitances connected to node $0$, including the capacitance to ground, the coupling capacitance $C_c$, and the capacitances between node $0$ and all other $M$ nodes of the quantum circuit. The vector $\vec{C}_V$ 
consists of the capacitances between node $0$ and each other node $1,2,..., M$, superscript $T$ denotes the transpose, and the $M\times M$ matrix $\boldsymbol{C}_M$ 
accounts for the capacitances of and between the $M$ nodes of the circuit not connected to the QCR.

The inverse of matrix $\boldsymbol{C}$ can be expressed as
\begin{equation}
\label{eqn:Cinv}
\boldsymbol{C}^{-1} = \left( {\begin{array}{*{20}{c}}
\frac{1}{C_N}+\frac{\alpha^2}{\tilde{C_0}} &   \frac{\alpha}{\tilde{C_0}} & \alpha \left(\tilde{C}_V^{-1}\right)^T  \\
\frac{\alpha}{\tilde{C_0}} & \frac{1}{\tilde{C_0}} & \left(\tilde{C}_V^{-1}\right)^T  \\
\alpha \tilde{C}_V^{-1} &  \tilde{C}_V^{-1}  & \tilde{\boldsymbol{C}}^{-1}_M \\
\end{array}} \right),
\end{equation}
where $\alpha$ is defined after Eq.~\eqref{eqn:HC}, and the scalar $\tilde{C}_0$, vector $\tilde{C}_V^{-1}$, and matrix $\tilde{\boldsymbol{C}}_M^{-1}$ are determined by the requirement $\boldsymbol{C} \boldsymbol{C}^{-1} = \boldsymbol{1}$ to be
\begin{eqnarray}
\tilde{C}_0 &=& C_0 -\alpha C_c -\vec{C}_V^T \boldsymbol{C}_M^{-1} \vec{C}_V, \\
\tilde{C}_V^{-1} &=& -\frac{1}{\tilde{C}_0} \boldsymbol{C}_M^{-1} \vec{C}_V, \\
\tilde{\boldsymbol{C}}_M^{-1} &=& \boldsymbol{C}_M^{-1} + \frac{1}{\tilde{C}_0} \left(\boldsymbol{C}_M^{-1} \vec{C}_V\right)\left(\boldsymbol{C}_M^{-1} \vec{C}_V\right)^T,
\end{eqnarray}
where the last product is to be treated as the outer product between the two vectors.

As an example, we calculate the inverse of the capacitance matrix for the QCR-transmon-resonator system described in Fig.~\ref{fig:07} and in Eq.~(\ref{eqn:H_QCR-tr-res}). In the notation used there, the capacitance matrix is given by
\begin{equation}
\boldsymbol{C} = \left( {\begin{array}{*{20}{c}}
   C_N             & -C_c                & 0  \\
   -C_c            & C_c+C_{\Sigma_\textrm{tr}}+C_g     & -C_g  \\
   0               & -C_g                & C_g+C \\
\end{array}} \right),
\end{equation}
where $C_{\Sigma_\textrm{tr}}=C_B+C_J$ is the total capacitance of the bare transmon. Using the equations above and the relation $C_N = C_\Sigma+C_c$, the effective capacitances entering the Hamiltonian are
\begin{align}
\tilde{C}_0 &= C_c+C_{\Sigma_\textrm{tr}}+C_g-\frac{C_c^2}{C_N}-\frac{C_g^2}{C_g+C}=\frac{\mathrm{det}(\boldsymbol{C})}{C_N(C_g+C)}, \notag\\
\tilde{C}_M &=\frac{\mathrm{det}(\boldsymbol{C})}{C_{\Sigma}(C_c+C_{\Sigma_\textrm{tr}}+C_g)+C_c(C_{\Sigma_\textrm{tr}}+C_g)}, \notag \\
\kappa_1&=\tilde{C}_V^{-1}=\frac{C_N C_g}{\mathrm{det}(\boldsymbol{C})},
\end{align}
where
\begin{eqnarray}
\mathrm{det}(\boldsymbol{C})&=&C_g\left[ C_{\Sigma}(C_c+C_{\Sigma_\textrm{tr}})+C_c C_{\Sigma_\textrm{tr}}\right] \\
&+&C\left[(C_{\Sigma}+C_c)C_g+C_{\Sigma}(C_c+C_{\Sigma_\textrm{tr}})+C_c C_{\Sigma_\textrm{tr}}\right]\notag
\end{eqnarray}
is the determinant of the matrix $\boldsymbol{C}$.

\section{Off and on regimes}
\label{app:onoff}

\begin{figure}[tb]
  \centering
  \includegraphics[width=1\linewidth]{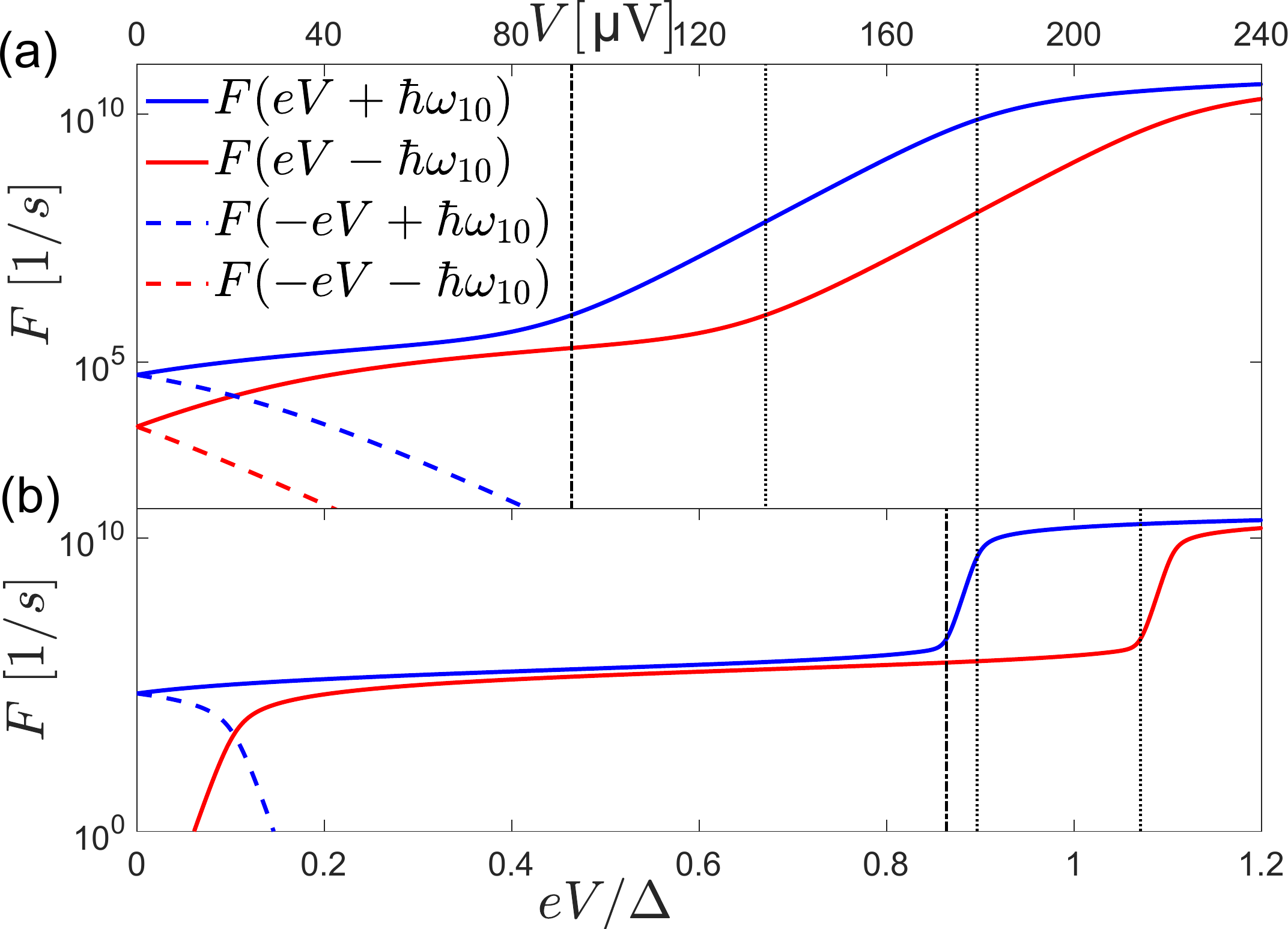}
  \caption{\label{fig:10} Shifted normalized tunneling rate functions $F(eV\pm \hbar \omega_{10})$ as functions of the bias voltage $V$ at the electron temperature (a)~$k_B T_N/\Delta = 0.043$ and (b)~$k_B T_N/\Delta = 0.0043$. The vertical dash-dotted lines at $eV=E_\mathrm{co}-\hbar \omega_{10}$ bound the off-state regime from above.  The region between the vertical dotted lines corresponds to low effective temperature. In panel (a), where $T_N> T_N^\mathrm{co}$, the vertical dotted lines are at voltages $eV=E_{\rm co}+\hbar \omega_{10}$ and $eV=\Delta-\hbar \omega_{10}$. In panel (b), where $T_N< T_N^\mathrm{co}$, the vertical dotted lines are at $eV=\Delta-\hbar \omega_{10}$ and $eV=E_{\rm co}+\hbar \omega_{10}$. The Dynes parameter is $\gamma_D = 10^{-5}$ and the shift frequency is $\hbar \omega_{10}/\Delta=0.1$.}
\end{figure}

In this appendix, we consider in detail how to determine the bias voltages defining the off- and on-states of the QCR and to calculate the corresponding relaxation rates. We begin with the off-state which we define as the bias region, in which the transition rates $\Gamma_{10}$ and $\Gamma_{01}$ are sufficiently small, such that the effect of the QCR on the quantum circuit can be neglected. According to Eq.~(\ref{eqn:Gamma}), we have $\Gamma_{10} \propto F(eV+\hbar \omega_{10})+F(-eV+\hbar \omega_{10})>\Gamma_{01} \propto F(eV-\hbar \omega_{10})+F(-eV-\hbar \omega_{10})$, where we neglect the small energy shift by $E_N$ (see Sec.~\ref{sec:circuitcoupling}) and use the fact that $F(E)$ is a monotonically increasing function (see Appendix~\ref{app:F_E}). Thus, up to a factor of 2 and to the prefactors from Eq.~(\ref{eqn:Gamma}), we can use $F(eV+\hbar \omega_{10})$ as an upper bound for $\Gamma_{10}$. Since we have shown in Appendix~\ref{app:F_E} that $F(E)$ increases slowly up to energy $E_\mathrm{co}$ and increases exponentially after that, the upper voltage for the off-state is approximately given by $eV = E_\mathrm{co}-\hbar \omega_{10}$. The off-state relaxation rate $\Gamma^{\rm off}_{10}$ for a transmon is given in Eq.~(\ref{gamma_off}).
Our estimate is confirmed by numerical calculations, as shown in Fig.~\ref{fig:10} for two different temperatures. At $eV=E_{\rm co}-\hbar \omega_{10}$, we find $F(eV \pm \hbar \omega_{10})\gg F(-eV \pm \hbar \omega_{10})$ and that the increase in $F$ compared to zero bias is less than one order of magnitude.

Let us consider the on-state, in which we aim the relaxation rate to be as high as possible while maintaining a low effective temperature $T$. Based on the discussion of the off-state, in the on-state we have $eV>E_\mathrm{co}-\hbar \omega_{10}$, and hence $\Gamma_{10} \propto F(eV+\hbar \omega_{10})$, $\Gamma_{01} \propto F(eV-\hbar \omega_{10})$. Therefore for the effective temperature we have $T \propto 1/\ln[F(eV+\hbar \omega_{10})/F(eV-\hbar \omega_{10})]$. Naively, for the relaxation rate to be high we require $eV+\hbar \omega_{10} \gtrsim \Delta$, but for temperature to be low we want $F(eV-\hbar \omega_{10})$ small, or equivalently $eV-\hbar \omega_{10} \lesssim E_\mathrm{co}$. These two condition together define the voltage range for low effective temperature $\Delta - \hbar \omega_{10}\lesssim eV \lesssim E_\mathrm{co} + \hbar \omega_{10}$. However, the two conditions can be satisfied simulataneously only if $E_\mathrm{co} \gtrsim \Delta - 2\hbar \omega_{10}$. Using Eq.~(\ref{eqn:Eco}), we find that this condition is equivalent, up to the leading order, to $T_N \lesssim T_N^\mathrm{co}$. Hence, in this low-temperature regime, the operation point, i.e., the maximum voltage satisfying our requirements, is $eV = E_\mathrm{co}+\hbar \omega_{10}$, which is identical to Eq.~(\ref{eq:eVmax}). At this voltage, the effective temperature is approximately $T^\mathrm{min}(0)$ [Eq.~(\ref{eqn:T^min_T})] and the relaxation rate can be estimated by substituting Eq.~(\ref{eqn:Fabovegap}) into Eq.~(\ref{eqn:Gamma}), where the contribution from $F(-eV+\hbar \omega_{10})$ can be neglected. For the transmon, this results in the transition rate in Eq.~(\ref{eq:GammaonlowT}). The low-temperature regime is exemplified in Fig.~\ref{fig:10}(b).

The simple argument above fails for $T_N > T_N^\mathrm{co}$. In this case, the low-effective-temperature region is defined by $E_\mathrm{co}+\hbar \omega_{10} \lesssim eV \lesssim \Delta - \hbar \omega_{10}$. At such voltages, both $F(eV+\hbar \omega_{10})$ and $F(eV-\hbar \omega_{10})$ are approximately given by the thermal activation formula of Eq.~(\ref{eqn:Fthermal}), implying $T \approx T_N/2$. 
The operation point is clearly $eV=\Delta-\hbar \omega_{10}$. At this voltage, for the relaxation rate of Eq.~(\ref{eqn:Gamma}) we have
\begin{eqnarray}
\Gamma^{\rm on}_{10} & \simeq &
\left|M_{10}\right|^2 \frac{2R_{K}}{R_T} F(\Delta) \\
& \approx & \bar \gamma\frac{\pi}{\omega_{10}} \frac{1}{h}\sqrt{\frac{k_B T_N\Delta}{2}} \times 1.072
\approx 0.38\times \bar \gamma \frac{\sqrt{k_B T_N\Delta}}{\hbar \omega_{10}},\notag
\end{eqnarray}
where on the second line we have used Eq.~(\ref{eqn:Fdelta}) and the leading-order form of the matrix element  for the transmon, Eq.~(\ref{eqn:Mtr}).

\section{On-state dephasing}
\label{app:on_deph}

In this appendix, we briefly consider dephasing by the QCR in the on-state. For the temperature region where the normal-metal temperature is above the cross-over value $T_N > T_N^{\rm co}$ and in the thermal activation regime for the bias voltage, the rate of elastic transitions obtained from Eq.~\eqref{eqn:elas} raises exponentially as the bias voltage approaches the superconducting gap edge
\begin{equation}
    \Gamma_{\rm el}^{\rm on}(eV)\approx\frac{R_{K}}{R_{T}}  \frac{\sqrt{2 \pi k_{\rm B}T_{N} \Delta}}{h }\exp\left(\frac{eV-\Delta}{k_{\rm B}T_{N}}\right).
\end{equation}
At the operation voltage $eV=\Delta-\hbar \omega_{10}$, the jump rate, that is approximated from the elastic transition rate, is given by
\begin{equation}
    \chi^{\rm on}\approx\frac{R_{K}}{R_{T}} \frac{\sqrt{2 \pi k_{B}T_{N} \Delta}}{h}\exp\left(-\frac{\hbar \omega_{10}}{k_{B}T_{N}}\right).
\end{equation}
In the low-temperature regime, for $T_N<T_N^{\rm co}$, the bias point is above the superconductor gap, and hence the elastic rate can be expressed by Eq.~\eqref{eq:F_0T} as
\begin{equation}
      \chi^{\rm on}=\Gamma_{\rm el}^{\rm on}(eV_{\rm max})= \frac{2R_{K}}{R_{T}}\frac{\sqrt{(eV_{\rm max})^2-\Delta^2}}{h}
\end{equation}
where $V_{\rm max}$ is the on-state voltage defined in Eqs.~\eqref{eq:eVmax} and~\eqref{eq:eVmax2}.

Using $T_{N}=$~\SI{100}{\milli\kelvin} for the thermal activation regime and $T_{N}=$~\SI{10}{\milli\kelvin}$<T_{N}^{\rm co}$ for the low-temperature regime, we estimate the jump rates in the on-state to be $\chi^{\rm on}/2\pi=180$~MHz (thermal activation) and $\chi^{\rm on}/2\pi=3$~GHz (low temperature), see Fig.~\ref{fig:04}. With $E_{J}/E_{C}=$~\numrange{50}{100} and other parameters from Table~\ref{tab:parameters}, the transmon charge dispersion~\cite{Koch07, Schreier08} between the two lowest levels results in a jump amplitude $\xi$ of the order $\xi/2\pi\approx$~\SI{3}{\hertz}--\SI{6}{\kilo\hertz}. In the on-state, the jump frequency $\chi^{\rm on}$ is in the~\SI{}{\giga\hertz}-scale, which is much greater than even the largest jump amplitude $\xi$. Therefore, the additional dephasing rate is in the fast-jumping regime $\xi^2/4\pi \chi \ll $~\SI{1}{\hertz}, which is negligible in comparison to the typical dephasing rates of transmons $\Gamma^b_2/2\pi\approx$~\SIrange{1}{4}{\kilo\hertz} where $\Gamma_2^b$ is the bare qubit dephasing rate in the absence of the QCR~\cite{Paik11, Rigetti_2012, Chang_2013, Barends13}.

\bibliography{references}{}
\end{document}